\documentclass[fleqn,usenatbib]{mnras}

\usepackage{newtxtext,newtxmath}

\usepackage[T1]{fontenc}
\usepackage{xcolor}
\usepackage[utf8]{inputenc}
\usepackage{CJKutf8}

\DeclareRobustCommand{\VAN}[3]{#2}
\let\VANthebibliography\thebibliography
\def\thebibliography{\DeclareRobustCommand{\VAN}[3]{##3}\VANthebibliography}

\author[W. Zhang et al.]{
W. Zhang~
\begin{CJK*}{UTF8}{gbsn}
(张文达)
\end{CJK*}
$^{1,2}$\thanks{wdzhang@nao.cas.cn},
I. E. Papadakis$^{3,4}$,
M. Dov\v{c}iak$^2$,
M. Bursa$^2$,
V. Karas$^2$
\\
$^1$National Astronomical Observatories, Chinese Academy of Sciences, A20 Datun Road, Beijing 100101, China\\
$^2$Astronomical Institute of the Czech Academy of Sciences, Bo{\v c}n{\'i} II 1401, CZ-14100 Prague, Czech Republic\\
$^3$Department of Physics and Institute of Theoretical and Computational Physics, University of Crete, 71003 Heraklion, Greece\\
$^4$Institute of Astrophysics, FORTH, GR-71110 Heraklion, Greece}

\usepackage{graphicx}	

\newcommand\codename{\textsc{Monk}}

\newcommand\Rg{R_{\rm g}}
\newcommand\Tg{T_{\rm g}}
\newcommand\Rc{R_{\rm c}}
\newcommand\Te{kT_{\rm e}}
\newcommand\taut{\tau_{\rm T}}
\newcommand\nubtl{\nu_{\rm btl}}
\newcommand\mdot{\dot{m}_{\rm Edd}}
\newcommand\mbh{M_{\rm BH}}
\graphicspath{{./figures/}}
\defcitealias{epitropakis_x-ray_2017}{EP17}

\title[Compton lags]{A theoretical study of the time-lags due to Comptonization and the constraints on the X-ray corona in AGNs}

\date{Accepted XXX. Received YYY; in original form ZZZ}

\pubyear{2022}

\begin{document}
\label{firstpage}
\pagerange{\pageref{firstpage}--\pageref{lastpage}}
\maketitle

\begin{abstract}
We study the Fourier time-lags due to the Comptonization of disc-emitted photons in a spherical, uniform, and stationary X--ray corona, which located on the rotational axis of the black hole. We use \textsc{Monk}, a general relativistic Monte-Carlo radiative transfer code, to calculate Compton scattering of photons emitted by a thin disc with a Novikov-Thorne temperature profile. We find that the model time-lags due to Comptonization remain constant up to a characteristic frequency and then rapidly decrease to zero at higher frequencies. We provide equations which can be used to determine the time-lags and cross spectra for a wide range of values for the corona radius, temperature, optical depth, height, and for various accretion rates and black hole masses. We also provide an equation for the X--ray luminosity of a single corona, as a function of the its characteristics and location above the disc. Remarkably, the observed X--ray time-lags of nearby, bright active galaxies can be successfully reproduced by inverse Comptonization process of multiple dynamic coronae.

\end{abstract}


\section{Introduction}

Active Galactic Nuclei (AGNs) are the most luminous, persistent sources in the Universe. A substantial part, of the order of $\sim 1-10$\%, of their bolometric luminosity is emitted in the 2--10 keV band \citep[e.g.][]{lusso_bolometric_2012}, and the observed X--ray variations are of high amplitude, and very fast. These properties suggest that X--rays are produced in a small region, close to the central source, where most of the accretion power is released. According to the current paradigm, X-rays are thought to be produced by Comptonization of optical/UV photons, emitted by the disc, by hot electrons  (with a temperature of $\sim 10^9$ K), in a region which is known as the ‘X-ray corona’. The main evidence for this comes from the study of the energy spectra of many X--ray bright AGNs, which are well described by a power-law continuum, with a photon index of $1.5\le \Gamma \le 2.5$, and an exponential cut-off at energies of the order of tens to hundreds of keV \citep[e.g.][]{lubinski_comprehensive_2016,ricci_bat_2017,tortosa_nustar_2018,panagiotou_nustar_2020}. However, the geometrical properties of the X-ray corona are still not fully understood.

Numerous timing studies of a few, X--ray bright, highly variable AGNs have detected continuum, `hard' time-lags \citep[i.e.\ the high energy, `hard', X--ray photons lag behind the lower energy, 'soft', photons;][]{papadakis_frequency-dependent_2001,mchardy_combined_2004,arevalo_spectral-timing_2006,arevalo_x-ray_2008,sriram_energy-dependent_2009,epitropakis_x-ray_2017,papadakis_long_2019}. In general, the hard continuum time-lag has a power-law dependence on Fourier frequency (with a slope close to $-1$) and their amplitude is proportional to the logarithm of the energy separation between the energy bands photon energy. Continuum hard lags with the same properties have also been detected in black-hole X-ray binaries, which are the stellar-mass counterparts of AGNs \citep[e.g.][]{miyamoto_delayed_1988,miyamoto_x-ray_1989,nowak_phase_1996,nowak_rossi_1999,cui_phase_2000,crary_hard_1998,reig_energy_2003,altamirano_evolution_2015}.

Comptonization predicts delays between the higher and lower energy X--ray photons.  When traveling inside the hot X-ray corona, the photons scatter off the hot electrons. Photons gain energy after each scattering and, on average, higher energy photons are scattered more times. This effect should naturally give rise to hard lags. However, it was soon realized that the Comptonization process cannot explain the observed hard lags, as it predicts a constant,  frequency-independent time-lag \citep{miyamoto_delayed_1988}. Later \citet{kazanas_temporal_1997} and \citet{hua_phase_1997} studied the temporal properties of the Comptonized radiation using Monte-Carlo methods. They investigated how the time-lags depend on the optical depth as well as the density profile of the corona. They found that the observed time-lags can be explained if the X-ray corona is extended and inhomogeneous, with the electron density decreasing roughly inversely proportional to the distance from the central source.  

We study the time-lags due to Comptonization in the case of a spherical, homogeneous and isothermal corona, located above the supermassive black hole (BH). This is the so-called lamp-post geometry, which has been considered widely in the literature in the past. This geometry can explain the relativistic broadening of the iron-line at 6.4 keV, as well as the delays in the observed optical/UV variations in many AGNs. We study in detail the dependence of the Comptonization time-lags on the physical properties of the corona, on the accretion rate and on the BH mass, in the case of a stationary corona, which responds to disc photon variations, but also in the more general case of multiple coronae, which are created at various heights, have various size, temperature, optical depth, and duration. Our main objective is to put constraints to the properties of the X-ray corona, using the results from recent studies of the X--ray continuum time-lags in AGNs. If Comptonization of optical/UV photons is the mechanism responsible for the X-ray emission in AGNs, time delays due to Comptonization must be present in the observed light curves at different energy bands. Even if the time-lags due to Comptonization cannot explain the observations, they should at least be consistent with the observed time-lags.

In \S 2 we discuss how we compute the corona response to the photons emitted by the disc, using a modified version of \codename{}, a Monte-Carlo radiative transfer code that includes all general relativistic effects \citep{zhang_constraining_2019}. In \S 3 we present the results from a detailed study of the Comptonization time-lags and the corresponding cross spectra, as well as the formalism for computing the time-lags of multiple dynamic coronae. In \S 4 we compare the model predictions with observations of nearby, bright AGNs, as reported by \citet[EP17 hereafter]{epitropakis_x-ray_2017}. In \S 5 we summarize and discuss our results.

\section{The model set-up}
\subsection{The disc/corona geometry}

We assume a razor-thin accretion disc located on the equatorial plane around a rotating black hole, between the innermost stable circular orbit and $100~\Rg$ ($\Rg \equiv \rm {GM_{BH}/c^2}$, is the gravitational radius). We assume that $F_{\rm NT}(r)$, the flux density from one side of the disc as measured by an observer corotating with the disc, follows the Novikov-Thorne temperature profile \citep[][while in practice we use eq. 11b of the latter]{novikov_astrophysics_1973,page_disk-accretion_1974}. The accretion disc emits locally with color-corrected blackbody spectrum, with effective temperature $T_{\rm eff}(r)\equiv [F_{\rm NT}(r)/\sigma_{\rm SB}]^{1/4}$, and a color correction factor of 2.4 \citep{ross_spectra_1992}, where $\sigma_{\rm SB}$ is the Stefan-Boltzmann constant. The emission of the disc is isotropic in the rest frame of the disc fluid. We also assume a spherical, homogeneous and isothermal X-ray corona, located on the black hole rotational axis. The corona is parametrized by the height above the equatorial plane, $h$, the radius, $\Rc$, the electron temperature, $\Te$, and the Thomson optical depth, $\taut \equiv n_e \sigma_{\rm T} \Rc$ ($n_e$ is the electron number density, and $\sigma_{\rm T}$ is the Thomson scattering cross section).

First we will study the time-lags in the case when a burst of UV/optical photons (emitted from the disc) enter the corona at the same time. We assume that the corona temperature and optical depth do not vary with time. The UV/optical photons will be Compton up-scattered to X-rays. Our objective is to compute the number of photons that will escape the corona as a function of time, in various energy bands (i.e.\ the energy dependent corona response to the input seed photons, due to Comptonization). We will use the resulting light curves to compute the cross-spectrum and time-lags between various energy bands. 

\subsection{The Comptonization code}

\codename{} is a Monte-Carlo radiative transfer code that includes all general relativistic (GR) effects \citep{zhang_constraining_2019}. Basically, \codename{} samples photons from several emission processes (blackbody, bremsstrahlung, synchrotron, phenomenological power-law, etc.), and ray-traces the emitted photons along null geodesics in the Kerr spacetime. If the photons are travelling through media, \codename{} also takes into account various interactions between the medium and the photons, including Compton scattering, bremsstrahlung self-absorption, and synchrotron self-absorption.

We use an updated version of \codename{}. Compared with \citet{zhang_constraining_2019}, the latest version is capable of tracing the $t-$ and $\phi-$ coordinates of the Boyer-Lindquist coordinates in addition to the $r-$, $\theta-$components and the affine parameter $\lambda$, following \citet{carter_global_1968} and \citet{dexter_fast_2009}:

\begin{eqnarray}
t &=& \lambda + 2\int^r r(r^2 + a^2 - al) \frac{dr}{\Delta \sqrt{R(r)}},\\
\phi &=& a\int^r (r^2 + a^2 - al) \frac{dr}{\Delta\sqrt{R(r)}} \\ \nonumber
& & + \int^\mu \left(\frac{l}{1-\mu^2} - a\right) \frac{d\mu}{M(\mu)}.
\end{eqnarray}
This enables \codename{} to handle time-dependent problems as well as axially-asymmetric geometries. We neglect bremsstrahlung and synchrotron self-absorption and the only source of opacity is the Compton scattering of the hot electrons in the corona. 

\codename{} utilises the superphoton method where we sample and propagate ``superphotons'' instead of single photons. Each superphoton is actually a photon package that consists of many photons with identical energy and momentum, and is assigned a statistical weight $w$. The weight $w$ has the physical meaning of the number of photons generated per unit time as measured in a distant observer’s frame \citep[for detail, see][]{zhang_constraining_2019}. We start with sampling seed superphotons from the accretion disc. Each seed superphoton is characterized by its initial four-position $x^\mu_0\equiv \{t_0, r_0, \theta_0, \phi_0\}$, its initial wave vector $k^\mu$, its energy measured by an observer at infinity $E_\infty$, and its statistical weight $w$.

We propagate the superphotons along null geodesics in the Kerr spacetime. If the superphotons travel inside the corona, we propagate the superphoton with a step size much smaller than the scattering mean free path, and for each step of propagation we covariantly evaluate the Compton scattering optical depth $d\tau$ and calculate the scattering probability. If the superphoton is scattered, we sample the direction and energy of the scattered superphoton. The propagation terminates if the photon hits the accretion disc, if it falls into the black hole event horizon or if it arrives at infinity. In practice the propagation terminates if the superphoton reaches a radius $r_{\rm max} = 5\times10^4~\Rg$, which is far beyond the disc-corona system. We record the four-positions $x^\mu_\infty \equiv \{t_\infty, r_{\rm max}, \theta_\infty, \phi_\infty\}$ and energies $E_\infty$ of the photons arriving at infinity. Interested readers are referred to Sec. 2.2, 2.5, and 2.6 of \citet{zhang_constraining_2019} for detailed descriptions of the treatment of seed photon sampling, photon propagation, and Comptonization in \codename, respectively.

\subsection{Computation of the corona response}

We assume an observer at infinity, located at a position with polar angle $\theta_{\rm obs}=30^\circ$, and without loss of generality azimuthal angle $\phi_{\rm obs} = 0^\circ$. The observed energy spectrum and the photon rate in an energy band (i.e.\ the ``light curve" in this band) are computed using photons that fall onto a pixel centered at $\{\theta_{\rm obs}, \phi_{\rm obs}\}$. The pixel spans $\Delta \theta$ in the polar angle and $\Delta \phi$ in the azimuth angle. The solid angle subtended by the pixel is therefore 

\begin{equation}
\Delta \Omega \equiv \Delta \phi \int_{\theta_{\rm obs} - \Delta \theta /2}^{\theta_{\rm obs} + \Delta \theta /2} {\rm sin}\theta \,{\rm d}\theta.
\end{equation}
\noindent
In this paper we take $\Delta \theta = 10^\circ$, and $\Delta \phi = 360^\circ$, which is appropriate for an axially symmetric corona. 

The corona response in the energy band between $E_{\rm min}$ and $E_{\rm max}$ $N(E,t_j)$, is defined as the number of photons detected by the observer per unit time, and is given by 

\begin{equation}
 N(E,t_j) = \frac{4\pi\sum_k w_k(t)}{\Delta \Omega \Delta t},
\end{equation}
\noindent
where $E$ is the mean of the energy band, and $\Delta t$ is the width of the time bin\footnote{We chose $\Delta t$ to be equal to $0.46875\,\Tg$, where $\Tg=GM_{\rm BH}/c^3$ is the light travel time for one gravitational radius. This is equal to 49.27 s for a 10$^7$M$_{\odot}$ BH.}. $\sum_k w_k(t)$ is the sum of all photons falling in the pixel with $E_\infty \in [E_{\rm min}, E_{\rm max})$, and $t_{\infty} \in [t_j,t_j+\Delta t)$ ($t_j=j\Delta t$, $j=0,1, \dots, N_{\rm lc} -1$; $N_{\rm lc}$ is the number of the light curve points). 
We study the corona response in the 1--2, 2--4, 4--6, 6--10 and 10--20 keV bands. Time-lags in these bands are well determined, both for X-ray binaries and AGNs, so we can meaningfully compare the model predictions with observations. 

\section{Computation of the cross-spectrum and time-lags due to thermal Comptonization}
\label{sec:timeanalysis}

\begin{table}
\begin{center}
\caption{Values of the corona/disc system that we consider.}
\label{tab:param}
\begin{tabular}{ll}
\hline
Parameter & Values \\
\hline
$\Rc(\Rg)$  & 1,2,5,{\color{red} 10} \\
$\Te$(keV)  & 50,{\color{red} 100},200 \\
$\taut$     & {\color{red} 0.5},1,2 \\
$h(\Rg)$    & 3,4,10,{\color{red} 20} \\
$\mdot$ $(\bar{E}_{\rm seed}$;keV)     & 0.01(0.036),{\color{red} 0.1(0.065)},1(0.12)\\
\hline
\end{tabular}
\end{center}
\end{table}

We consider a range of values for the physical parameters of the corona, i.e. $\Rc$, $\Te$, $\taut$, and $h$, as listed in Table \ref{tab:param}. In all cases, the BH mass is set to $10^{7}~\rm M_{\odot}$ and spin $a=0.998$. We also consider various accretion rates (listed in the bottom row of Table \ref{tab:param}, in $\dot{m}_{\rm Edd}$, the Eddington accretion rate; we define $\dot{m}_{\rm Edd}\equiv L_{\rm Edd}/\eta c^2$, where $L_{\rm Edd}$ is the Eddington luminosity, and $\eta$ is the radiative efficiency), in order to study the dependence of the Comptonization time-lags on the mean energy of the soft input photons (values in parenthesis in the bottom row). Numbers in red colour indicate the fiducial parameter value, when we investigate the dependence of the time-lags on other parameters.

For each parameter combination, we produce 20 lightcurves (i.e.\ corona responses), in the energy bands we consider. We use these light curves, and the \citet{epitropakis_statistical_2016} method to compute the model cross-spectrum (real and imaginary parts; see eqs. 15--20 in their Sec. 6), as well as the time-lags, between all energy bands and 2--4 keV. We define the 2--4 keV as the reference band because it is more representative of the X--ray continuum, with less ``contamination" from X--ray reflection features. The time-lags are defined in such a way so that positive time-lags imply that variations in the  2--4 keV band precede those in the other bands.

\begin{figure}
 \includegraphics[width=\columnwidth]{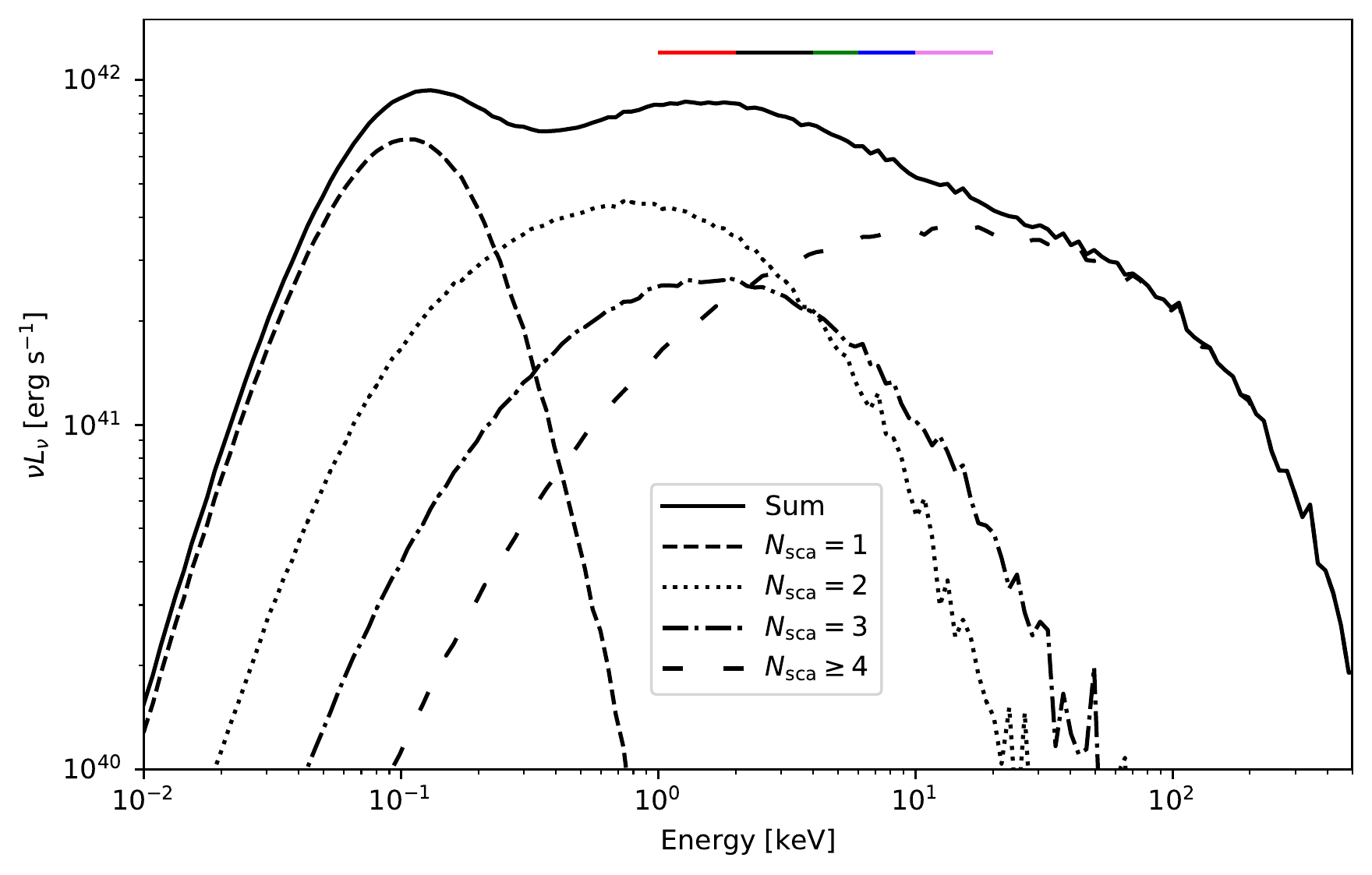}
 \caption{The energy spectrum emitted by an on-axis spherical corona, with $h=20~\Rg$, $\Rc=10~\Rg$, $\taut=0.5$, and $\Te=100~\rm keV$. The spectra of various scattering orders are plotted in different line styles, as indicated in the plot. The range of the energy bands for the computation of the time-lags is indicated by horizontal bars in different colors (red, black, green, blue and magenta, for the 1--2, 2--4, 4--6, 6--10, and 10--20 keV bands, respectively).}
 \label{fig:enspec_on}
\end{figure}

First we present the time-lags due to thermal Comptonization, and their dependence on their corona parameters. We will show later that, for most practical applications, it is the cross-spectrum that one needs to consider. Nevertheless, time-lags can help us understand the effects of thermal Comptonization on the delays between various energy bands.  

\subsection{An example}
\label{sec:example}

As an example, we discuss the time-lags in the case when $h=20~\Rg$, $\Rc=10~\Rg$, $\taut=0.5$, and $\Te=100~\rm keV$. They are typical of the time-lags in all cases we consider. Before discussing the timing properties, we present the energy spectrum in Fig.~\ref{fig:enspec_on}, together with the contribution of different scattering orders. The energy spectrum has a power-law shape above $\sim 2$ keV, with a slope of $\Gamma\sim 2.2$, and a high-energy cut-off of $\sim 200~\rm keV$. The dip in the spectrum is due to the fact that for lamp-post coronae, the seed photons direction is not isotropic, but coming from below the corona. In \codename{}, this effect can be taken into account as we compute exactly the direction and energy of the seed photons as they enter the corona.
This results in a weaker first-scattering component at smaller inclinations, and consequently a dip. This was also noticed by e.g.\ \citet{zhang_constraining_2019,haardt_anisotropic_1993}. It is dominated by photons which have been scattered $\bar{N}_{\rm sca}\sim 2$ and $\bar{N}_{\rm sca}\ge 4$ times, on average, in the 1--3 keV energy band and above $\sim 3$ keV, respectively ($\bar{N}_{\rm sca}(E)$ is the average number of scatterings of the photons at energy $E$).

\begin{figure}
 \includegraphics[width=\columnwidth]{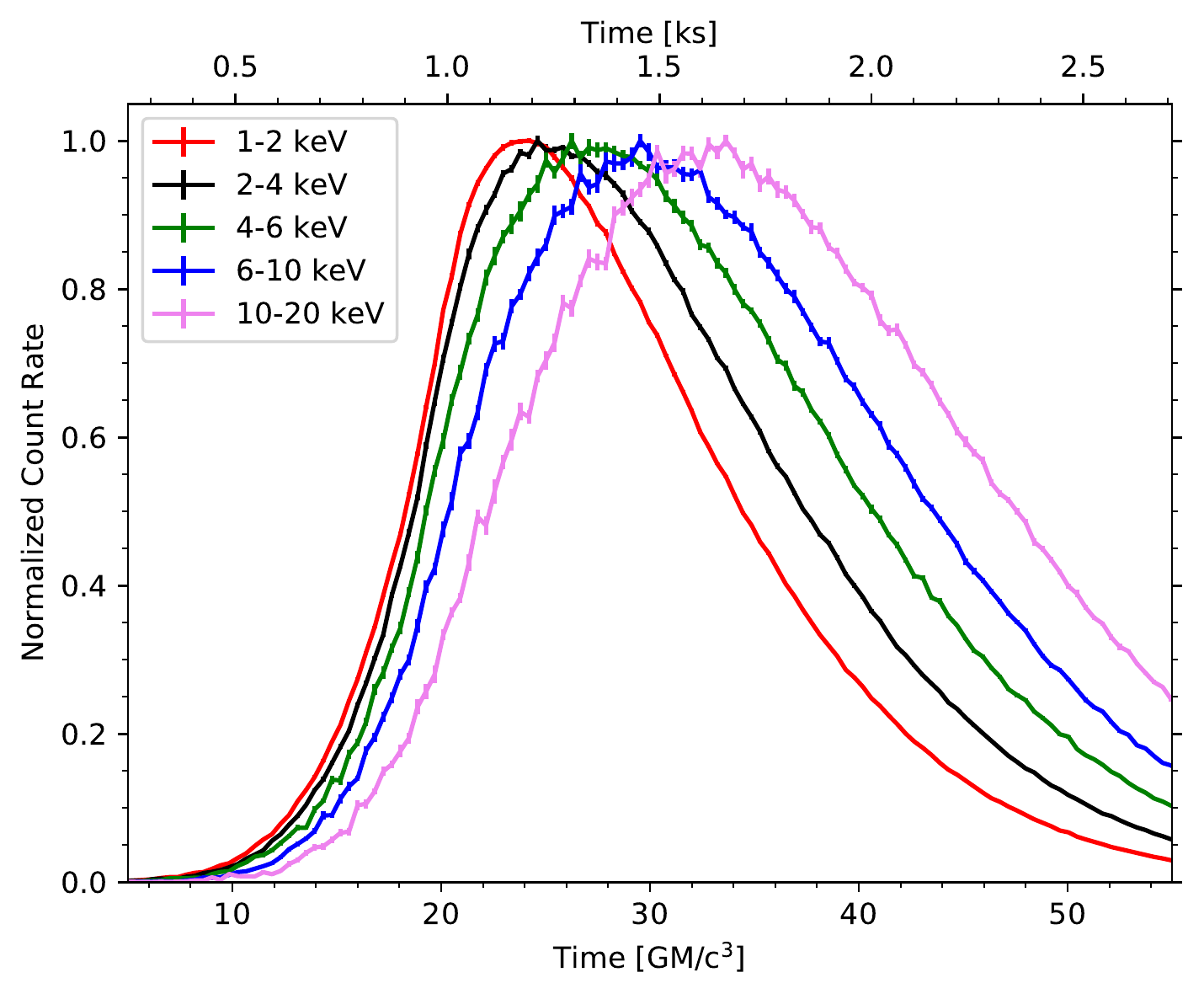}
 \caption{The response of the corona at different energy bands. The parameters of the disc-corona system are the same with Fig.~\ref{fig:enspec_on}. Time is measured in $\Tg(\equiv GM/c^3)$ and in kiloseconds in the bottom and upper $x-$axis, respectively. 
 \label{fig:on_lc}}
\end{figure}

Fig.~\ref{fig:on_lc} shows the mean of the 20 light curves in each band, which are representative of the energy dependent corona response. The shape of the response is the same in all bands: the X--ray flux increases at the beginning, reaches a peak, and then decreases slowly. The light curves are asymmetric about the peak, with fading tails longer than the rising parts. The response peak is delayed with increasing energy, since it takes more scatterings for the seed photons to reach higher energies.  

\begin{figure}
 \includegraphics[width=\columnwidth]{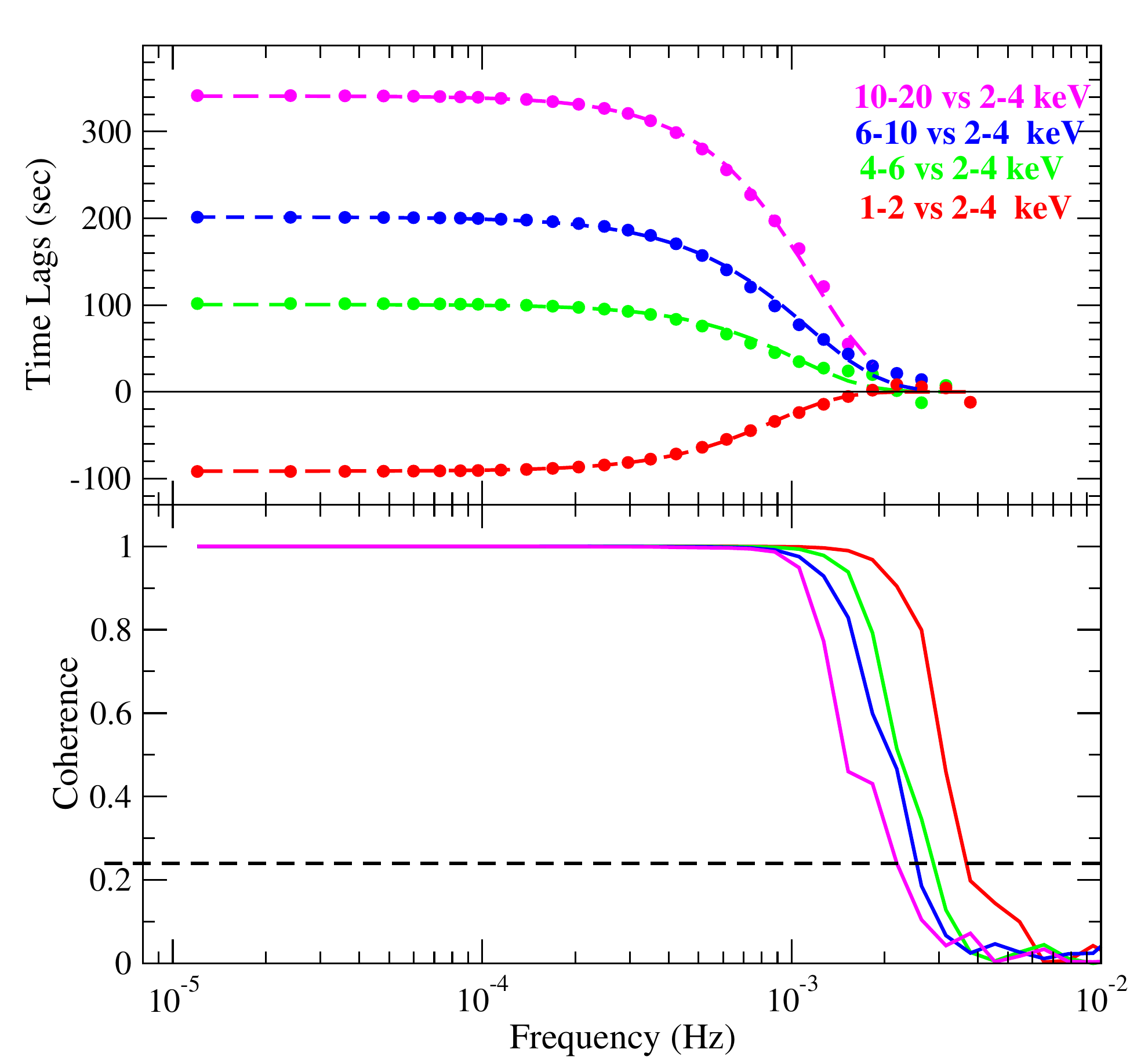}
 \caption{The time-lags, and the coherence of various energy bands with respect to the 2--4 keV band (top and bottom panels; see text for details). The dashed lines in the top panel show the best-fit models to the time-lags.
\label{fig:timing}}
\end{figure}

Fig.~\ref{fig:timing} shows the 1--2 vs 2--4, 4--6 vs 2--4, 6--10 vs 2--4, and 10--20 vs 2--4 time-lags and coherence, plotted with red, green, blue and magenta lines, respectively (in this, and all subsequent figures). The 1--2 vs 2--4 keV time-lags are negative because, as we mentioned, time-lags are computed with respect to the 2--4 keV band. All time-lags are flat at low frequencies and they decrease rapidly to zero at higher frequencies. 

The bottom panel in Fig.~\ref{fig:timing}  shows that coherence is equal to one at low frequencies and drops to zero at frequencies where the Poisson noise dominates the variations in the light curves\footnote{The Poisson noise is due to the fact that we have simulated a finite number of responses, with a finite number of seed photons.}. We do not study the coherence due to Comptonization in this work. We plot the model coherence in Fig.~\ref{fig:timing} to show that cross-spectra and time-lags (in all cases we study in this work) are computed at frequencies up to a maximum value (say $\nu_{\rm max}$), such that coherence($\nu_{\rm max})=1.2/(1+0.2m)$, where $m=20$, i.e. the number of light curves that we use to compute the time-lags. The horizontal dashed line in the bottom panel of Fig.~\ref{fig:timing} indicates the coherence at $\nu_{\rm max}$. According to \citet{epitropakis_statistical_2016}, cross-spectra and time-lags are unbiased at all frequencies up to $\nu_{\rm max}$. Therefore, the model cross-spectra and time-lags we compute  should be representative of the effects due to the Comptonization process.

To facilitate the discussion in the following sections of how the time-lags depend on the various corona parameters, we fit the time-lags with the model,
\begin{equation}
    \tau(E_{\rm i},E_{\rm ref},\nu)=A(E_{\rm i},E_{\rm ref}) e^{-[\nu/\nubtl(E_{\rm i},E_{\rm ref})]^2},
    \label{eq:model}
\end{equation}
\noindent
where $\nu$ is the variability frequency, $\rm i=1,2,3,4$, and $E_1$, $E_2, E_3,E_4,E_{\rm ref}$ are the mean photon energy of the 1--2, 4--6, 6--10, 10--20 keV and of the reference band, 2--4 keV, respectively. $A(E_{\rm i},E_{\rm ref})$ is the time-lags amplitude, i.e.\ the constant time-lag at low frequencies, while $\nubtl(E_{\rm i},E_{\rm ref})$ is the time-lags break frequency. The time-lags decrease above this frequency. The dashed lines in the top panel of Fig.~\ref{fig:timing} show the best-fit to the model time-lags. The function defined by eq.~\ref{eq:model} fits the model time-lags quite well, with the best-fit amplitude and $\nubtl$ being: ($-91.8~\rm sec$, $9\times10^{-4} ~\rm Hz$), ($101.7~\rm  sec$, $1\times10^{-3}~\rm Hz$), ($202~\rm sec$, $1.1\times10^{-3}~\rm Hz$), and ($339.7~\rm sec$, $1.2\times10^{-3}~\rm Hz$), in the case of the 1--2, 4--6, 6--10 and 10--20 vs 2--4 keV time-lags, respectively. The same function fitted similarly well the time-lags in all cases we considered in this work.

\begin{figure}
 \includegraphics[width=\columnwidth]{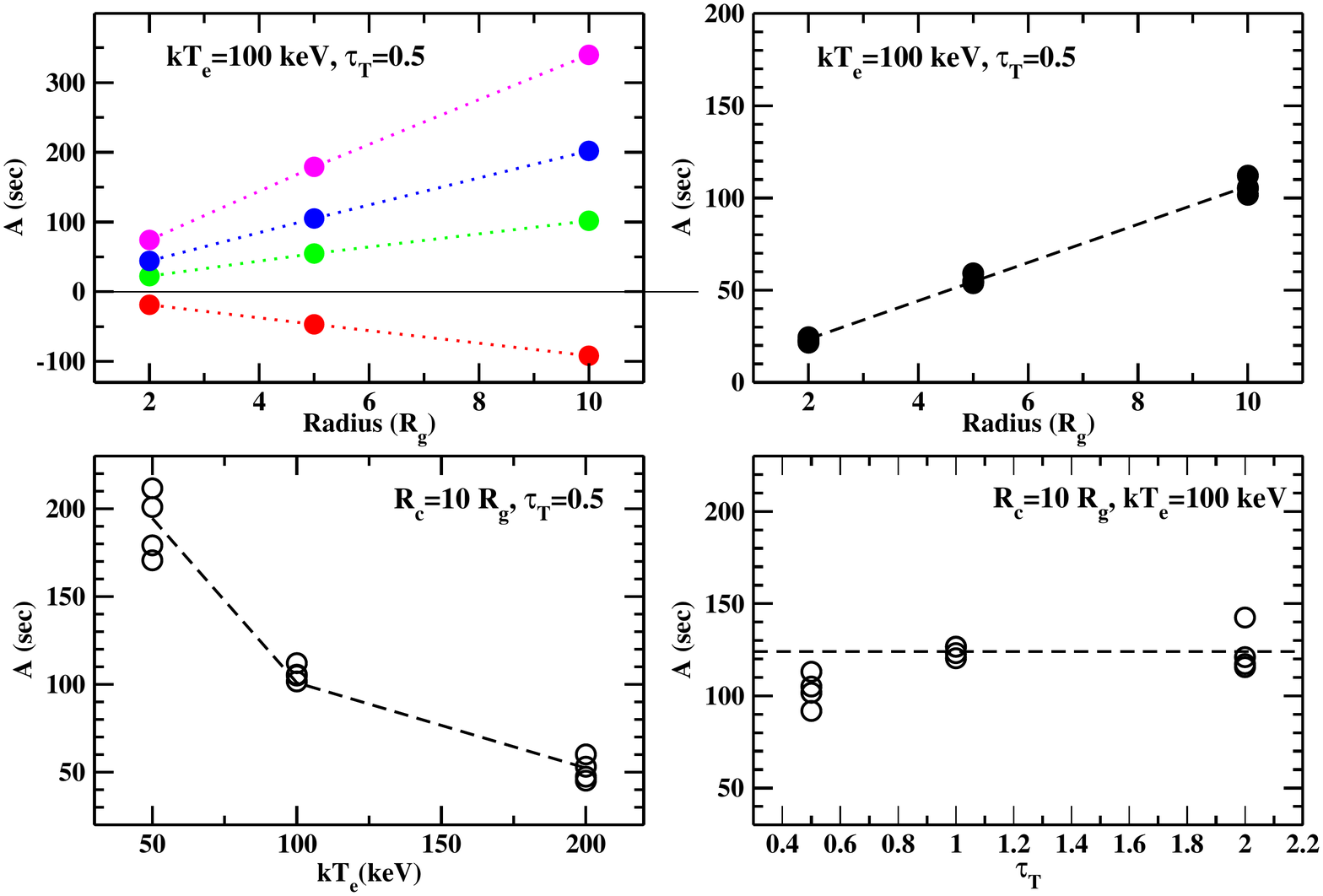}
\caption{The time-lags amplitude vs $\Rc$ ($h=20~\Rg$) and $\Te$, $\taut$ (upper and lower panels, respectively). The points in the upper-right panel and in the lower panels show the amplitude after the energy dependence is taken out. Values of the parameters that were kept constant are also shown. The dotted lines in the upper-left panel and dashed lines in other panels show the best-fit models.\label{fig:ampres}}
\end{figure}

\subsection{The time-lags amplitude}
\label{sec:ampres}

Fig.~\ref{fig:ampres} shows the time-lags amplitude plotted as a function of $\Rc$ (upper panels), $\Te$ and $\taut$ (lower panels). The upper-left panel shows the energy dependence of the amplitude (we remind that magenta, blue, green and red points show results for the 10--20, 6--10, 4--6, and 1--2 vs 2--4 keV time-lags). At fixed radius, $A(E_{\rm i},E_{\rm ref})$ increases with increasing energy separation, $\Delta E(E_{\rm i},E_{\rm ref})=|E_{\rm i}-E_{\rm ref}|$. We find that the amplitude increases proportional to $\log (E_{\rm i}/E_{\rm ref})$\footnote{We use the model energy spectra and we determined that: $\log (E_1/E_{\rm ref})=-0.27, \log (E_2/E_{\rm ref})=0.23, \log (E_3/E_{\rm ref})=0.42,$ and $\log (E_4/E_{\rm ref})=0.7$.}.

We multiply the 1--2, 6--10, and the 10--20 vs 2--4 keV time-lags with $0.23/\log(E_{\rm i}/E_{\rm ref} )$, for $i=1,3$ and 4, in order to normalize them to the 4--6 vs 2--4 keV time-lags. The best-fit parameters of the resulting time-lags are plotted in the upper-right panel of Fig.~\ref{fig:ampres}, and in the bottom panels of the same figure. Once the energy dependence is accounted for, the amplitude increases linearly with $\Rc$ (the dashed line in the upper-right panel of Fig.~\ref{fig:ampres} shows the best linear fit to the data). The bottom left panel in Fig.~\ref{fig:ampres} shows the time-lags amplitude vs $\Te$. The time-lags decrease roughly inversely proportional to the corona temperature  (the dashed line shows the best fit power-law model to the time-lags, where the slope is $\sim -1$). The bottom right panel in Fig.~\ref{fig:ampres} shows that the time-lags amplitude does not depend strongly on $\taut$ (the dashed line in this panel indicates the mean amplitude when $\taut=1$ and 2).

The results above can be explained as follows. Let us denote with $\bar{t}_{del}(E_{\rm seed},E_{\rm i})$ the average time for a seed photon to reach energy $E_{\rm i}$. We expect, 
\begin{equation}
\label{eq:tlagsamp}
 A(E_{\rm i},E_{\rm ref}) \propto  \Delta \bar{t} (E_{\rm i},E_{\rm ref})\simeq \Delta \bar{N}_{\rm sca}(E_{\rm i},E_{\rm ref})\times \bar{\lambda}/c, 
\end{equation} 

\begin{figure}
 \includegraphics[width=\columnwidth]{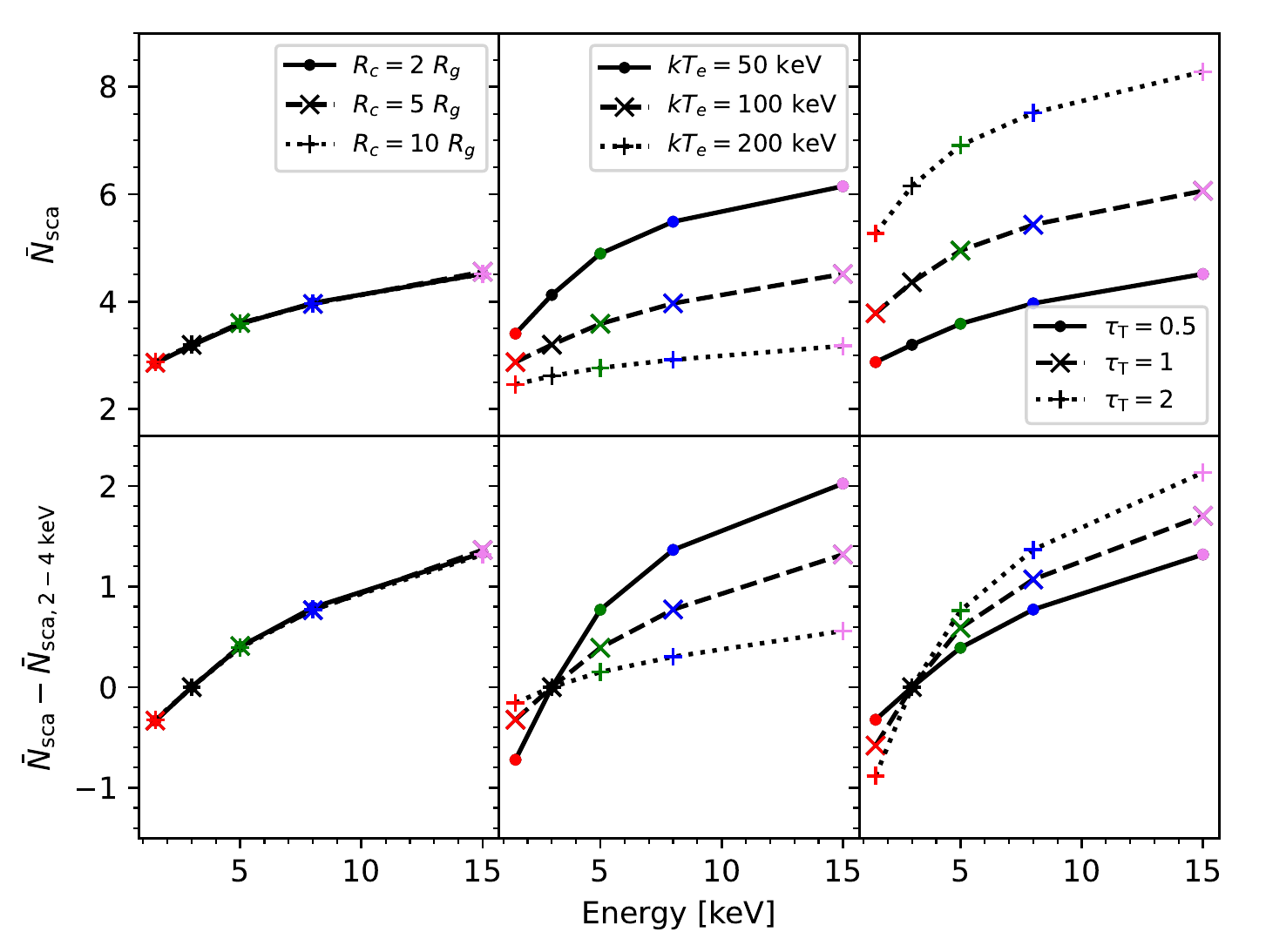}
\caption{{\it Top panels:}  $\bar{N}_{\rm sca}$ as function of energy for the three $R_{\rm c}$, $kT_{\rm e}$,, and $\tau_{\rm T}$
values that we consider (left, middle and right panel, respectively). {\it Bottom
panels}: The respective excess mean number of scattering with respect to
the 2–4 keV reference band, $\Delta \bar{N}_{\rm sca}$), for the same $R_{\rm c}$, $kT_{\rm e}$, and $\tau_{\rm T}$ values.}
\label{fig:scatterings}
\end{figure}

\noindent  where $\Delta \bar{t} (E_{\rm i},E_{\rm ref})= \bar{t}_{del}(E_{\rm seed}, E_{\rm i})-\bar{t}_{del}(E_{\rm seed},E_{\rm ref})$, $\Delta \bar{N}_{\rm sca}(E_{\rm i},E_{\rm ref})=\bar{N}_{\rm sca}(E_i)-\bar{N}_{\rm sca}(E_{\rm ref})$, and $\bar{\lambda}/c$ is the mean photon travel time between subsequent scatterings\footnote{We note that $\bar{\lambda}$ is only a fraction of the scattering mean free path, $\lambda_{T}\equiv 1/n_e \sigma_{T} \equiv \Rc/\taut$, which is comparable to the size of the corona (especially at low optical depths), but the photons have to travel a shorter distance, on average, in order to get scattered multiple times before escaping the corona.}. The upper and bottom panels in Fig.~\ref{fig:scatterings} show $\bar{N}_{\rm sca}(E)$ and $\Delta \bar{N}_{\rm sca}(E_i,E_{\rm ref})$ as a function of energy, for various corona parameters. The bottom panels show that the difference  between number of scatterings increases with energy, at all times. In fact, $\Delta \bar{N}_{\rm sca}(E_i,E_{\rm ref})$ increases with the logarithm of energy which thus explains the logarithmic dependence of time-lags on energy. The right, top and bottom panels show that $\bar{N}_{\rm sca}(E)$ and $\Delta \bar{N}_{\rm sca}(E_i,E_{\rm ref})$ do not depend on the corona radius. As long as the temperature is constant, $\bar{N}_{\rm sca}(E_i)$ is the same, irrespective of $\Rc$. Consequently,  $\Delta \bar{N}_{\rm sca}(E_i,E_{\rm ref})$ does not change with $R_c$. However, as the corona radius increases, with $\taut$ kept constant, the electron density $n_e$ decreases. Consequently, $\bar{\lambda}$ and the time-lags increase with increasing $\Rc$.

The mean energy the photon gains per scattering is \citep{pozdnyakov_comptonization_1983}:
\begin{center}
\begin{equation}
\frac{\overline{\Delta E}}{E} = \frac{4kT_e}{m_e c^2},
\label{eq:energy_gain}
\end{equation}
\end{center}

\noindent if $kT_e \gg E$ ($E$ is the original photon energy, and $m_e$ is the electron rest mass). The photon gains more energy per scattering as $kT_e$ increases. Consequently, both $\bar{N}_{\rm sca}(E_i)$ and $\Delta \bar{N}_{\rm sca}(E_i,E_{\rm ref})$ decrease with increasing $\Te$ (see middle panels in Fig.~\ref{fig:scatterings}). This explains why time-lags decrease with increasing corona temperature. 

The top right panel in Fig.~\ref{fig:scatterings} shows that $\bar{N}_{\rm scat}(E)$ increases with increasing $\taut$ in a non-linear way. This is because, as $\taut$ increases, the spectrum at energy $E$ is more and more dominated by the low-energy photons of higher scattering order than photons of a lower scattering order. Therefore, $\Delta \bar{N}_{\rm sca}(E_i,E_{\rm ref})$ also increases with increasing $\taut$ (bottom right panel in Fig.~\ref{fig:scatterings}). At the same time, since $\Rc$ is kept constant while $\taut$ increases, $n_e$ also increases, and hence $\bar{\lambda}$ decreases. These effects work in opposite directions (see eq.~\ref{eq:tlagsamp}), and for this reason, the time-lags amplitude is (almost) insensitive to $\taut$.

\begin{figure}
 \includegraphics[width=\columnwidth]{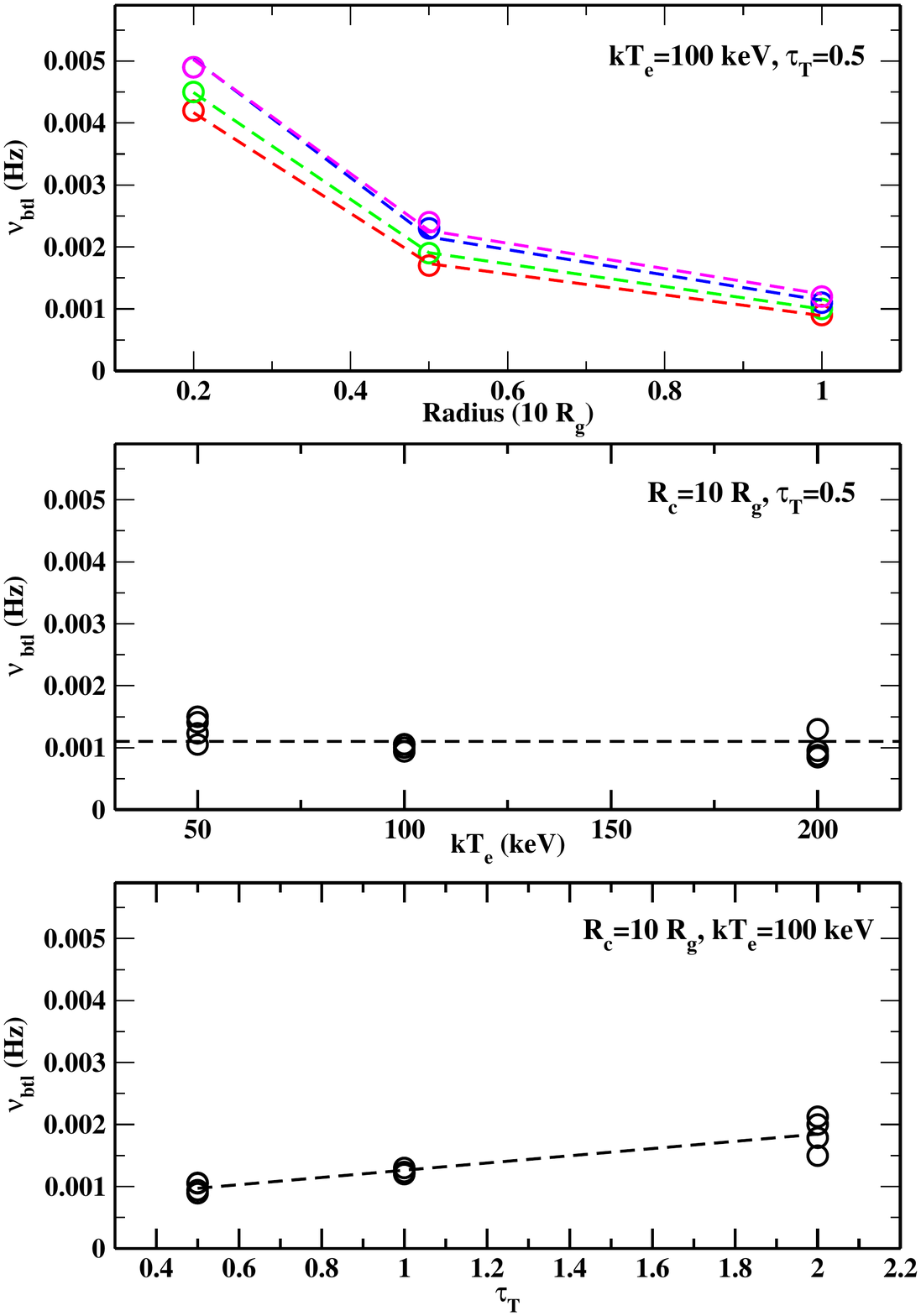}
 \caption{The time-lags break frequency, $\nubtl$, plotted as a function of $\Rc$, $\Te$ and $\taut$ (from top to bottom; $h=20~\Rg$). Red, green, blue and magenta points in the top panel show $\nubtl$ in the 1--2, 4--6, 6--10 and 10--20 vs 2--4 keV time-lags. The dependence of $\nubtl$ on energy is taken out in the other two panels. The dashed lines in the top and bottom panels show the best-fit models. Values of the parameters kept constant are shown in each panel.}
\label{fig:nubreakres}
\end{figure}

\subsection{The break frequency}
\label{sec:nubreakres}

Fig.~\ref{fig:nubreakres} shows a plot of the time-lags break frequency, $\nubtl$, as a function of $\Rc$, $\Te$ and $\taut$ (top, middle and bottom panels, respectively). The break frequency depends mainly on the corona radius: it decreases (almost) inversely proportional to $\Rc$ (the dashed lines in the top panel of Fig.~\ref{fig:nubreakres} show power-law model fits, with slope $\sim -1$). There is a positive correlation between $\nubtl$ and the separation between the energy bands, $\Delta (E_i,E_{\rm ref})$ (e.g.\ the magenta points are systematical above the red points in the top panel of Fig.~\ref{fig:nubreakres}). We find that both the amplitude, $N$, and the slope, $\beta$, of the $\nubtl - \Rc$ relation depend on energy: the power-law amplitude (slope) increases (decreases) linearly with $\log (E_i/E_{\rm ref}).$ When the dependence of $\nubtl$ on energy is taken out, $\nubtl$ does not depend on $kT_e$, while it is positively correlated with $\taut$ (middle and bottom panels in Fig.~\ref{fig:nubreakres}; the dashed lines in these plots indicate the mean $\nubtl$ and the best-fit linear model to the data, respectively). 

\begin{figure}
    \centering
    \includegraphics[width=\columnwidth]{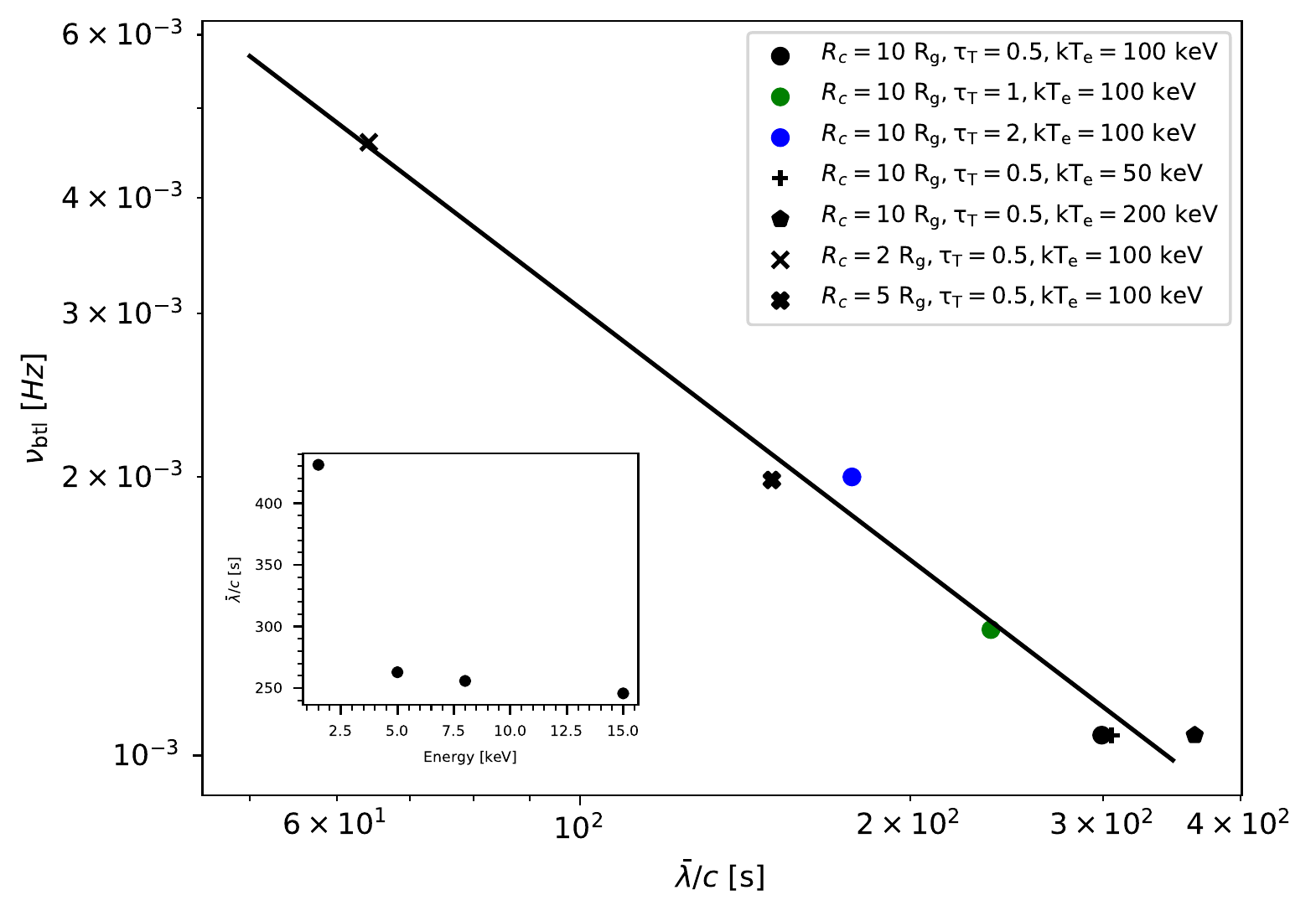}
    \caption{$\nubtl$ vs $\bar{\lambda}/c$ for various cases. The results are the average of all energy bands.
    The inset plot shows how $\bar{\lambda}/c$ depends on the photon energy.
    \label{fig:lagovernub_h20}}
\end{figure}

\noindent To explain the results above, we notice that $\nubtl$ should depend on some characteristic time scale of the corona and the scattering process. One natural choice is $\bar{\lambda}/c$, i.e.\ the mean photon travel time between subsequent scatterings. To test this hypothesis, we compute $\bar{\lambda}/c$ for all the parameters we considered, and we plot $\nubtl$ versus $\bar{\lambda}/c$ in  Fig.~\ref{fig:lagovernub_h20}. There is a strong anti-correlation between $\nubtl$ and $\bar{\lambda}/c$, which confirms that $\nubtl$ is mainly set by $\bar{\lambda}/c$. The solid line on this figure indicates the best-fit linear fit to the data (in the log-log space). The best fit slope is close to $-1$, which indicates that the break frequency is inversely proportional to the mean photon travel time.

As $\Rc$ decreases with $\taut$ fixed (or $\taut$ increases with $\Rc$ fixed), $n_e$ increases, leading to a smaller $\bar{\lambda}$, and hence to a higher $\nubtl$. Since $\bar{\lambda}$ does not depend on $\Te$\footnote{Strictly speaking, $\bar{\lambda}$ increases with $\Te$ for ultrarelativistic electrons as in this regime the Compton scattering cross-section decreases with $\Te$. However, in our work we are considering sub-relativistic plasma and this effect is negligible.}, $\nubtl$ should be independent of $\Te$, as we observe. Finally, the positive photon energy-$\nubtl$ correlation (shown in the upper panel of Fig.~\ref{fig:nubreakres}) is due to the fact that $\bar{\lambda}/c$ decreases with increasing photon energy, as shown in the inset plot in  Fig.~\ref{fig:lagovernub_h20} (for the case of a corona with $\Rc=10~ \Rg$, $\Te=100~\rm keV$, and $\taut=0.5$). This is a probabilistic effect: the photons detected at high energies have been scattered more times and must have encountered a smaller $\bar{\lambda}$, otherwise they would have escaped the corona before reaching the observed energy.


\subsection{The effects of the corona height}
\label{sec:height}

The time-lags depend on the corona height, $h$, due to its location with respect to the reprocessing accretion disc, path length along the ray through the extended corona from the initial point of the primary photon emission, and GR effects of light-bending and energy shift. To see how these effects modify the time-lags, we considered a corona with $\Rc=2~\Rg, \Te=100~\rm keV$, and $\taut=0.5$, located on the rotational axis, at height 4, 10 and 20 $\Rg$ above the BH ($\mdot=0.1$, M$_{\rm BH}=10^7$ M$_{\odot}$). We also considered a corona located at $h=3\Rg$, with a radius of $R_c=1~\Rg$ (a radius of $2~\Rg$ is not possible, as part of the corona would be inside the event horizon in this case). 

\begin{figure}
 \includegraphics[width=\columnwidth]{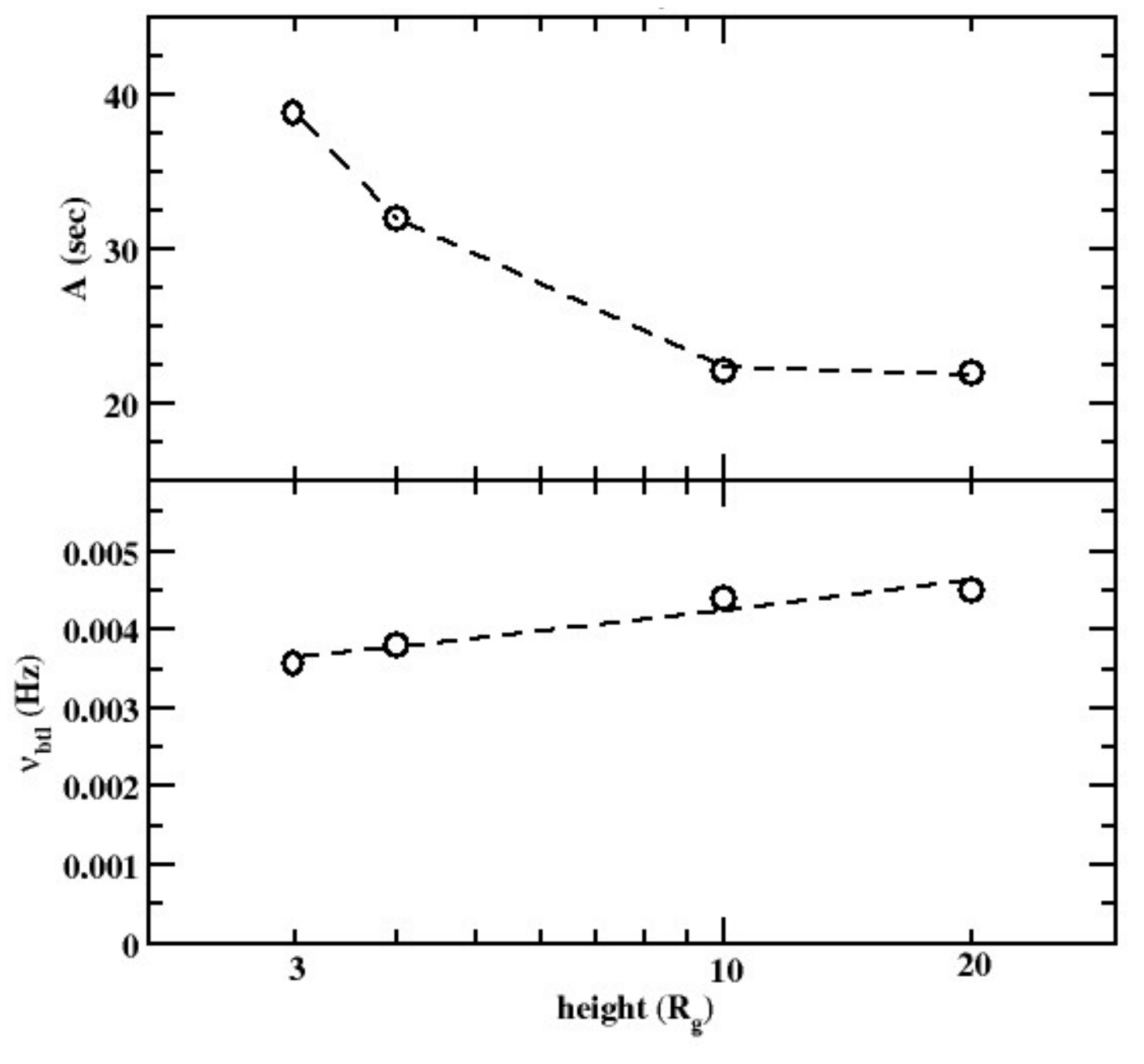}
 \caption{
 {\it Top and bottom panels:} The time-lags amplitude, $A$, and the break frequency, $\nubtl$, as a function of height. The dashed lines show the best-fit model. The y-axis range in this figure (as well as in Fig.~\ref{fig:mdotres}) is the same as in Fig.~\ref{fig:nubreakres}, to notice clearly which parameter mainly affects $\nubtl$.}
\label{fig:heightres}
\end{figure}

\begin{figure}
    \centering
    \includegraphics[width=\columnwidth]{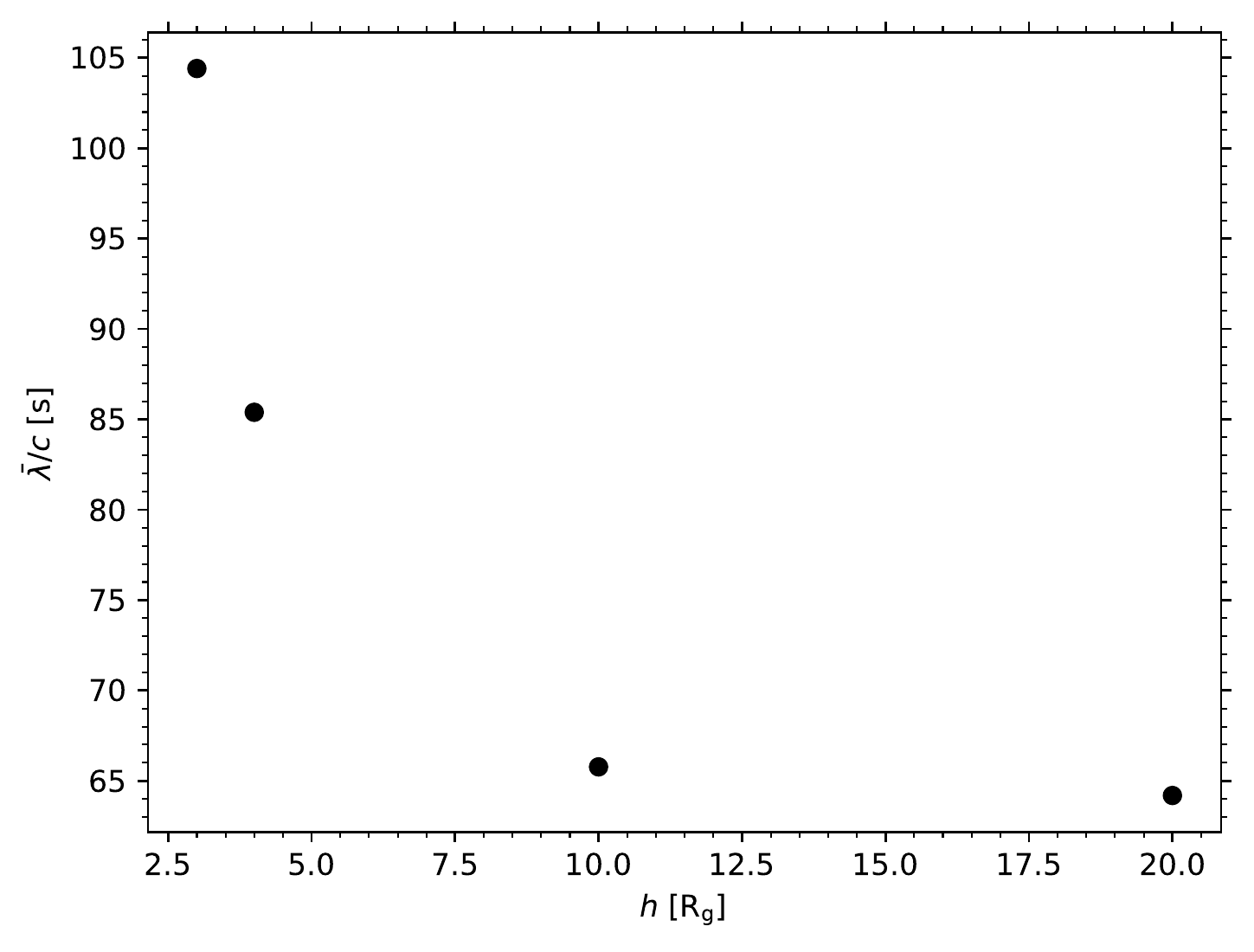}
    \caption{$\bar{\lambda}/c$ as a function of the corona height ($\Rc=2~\rm \Rg$, $\Te=100~\rm keV$, and $\taut=0.5$). For $h=3~\rm \Rg$ we measure the $\bar{\lambda}$ for the case of $\Rc=1~\rm \Rg$ and multiply the result with a factor of 2.
    \label{fig:lambdavsnub_height}}
\end{figure}

We found that the time-lags increase exponentially as we approach the BH, and $\nubtl$ increases with increasing corona height (top and bottom panel in Fig.~\ref{fig:heightres}; dashed lines show the best fits to the points). We believe that the increase of amplitude and the decrease of $\nubtl$ as we approach the BH is due to two reasons: 1. the relativistic time delay becomes longer as the corona approaches the black hole; 2. the same values of $\bar{\lambda}$, which is defined in the coordinate frame, correspond to longer proper distance at lower heights. We have shown that the mean light travel time between scatterings, $\bar{\lambda}/c$, affects both the time-lags amplitude (see eq.~\ref{eq:tlagsamp}), and the break frequency (see Fig.~\ref{fig:lagovernub_h20}). Fig.~\ref{fig:lambdavsnub_height} shows a plot of $\bar{\lambda}/c$ for the 4--6 keV band photons as a function of height. $\bar{\lambda}(4-6$ keV)$/c$ increases as we approach the BH. 

\subsection{The effects of the soft photon energy}
\label{sec:seeden}

\begin{figure}
 \includegraphics[width=\columnwidth]{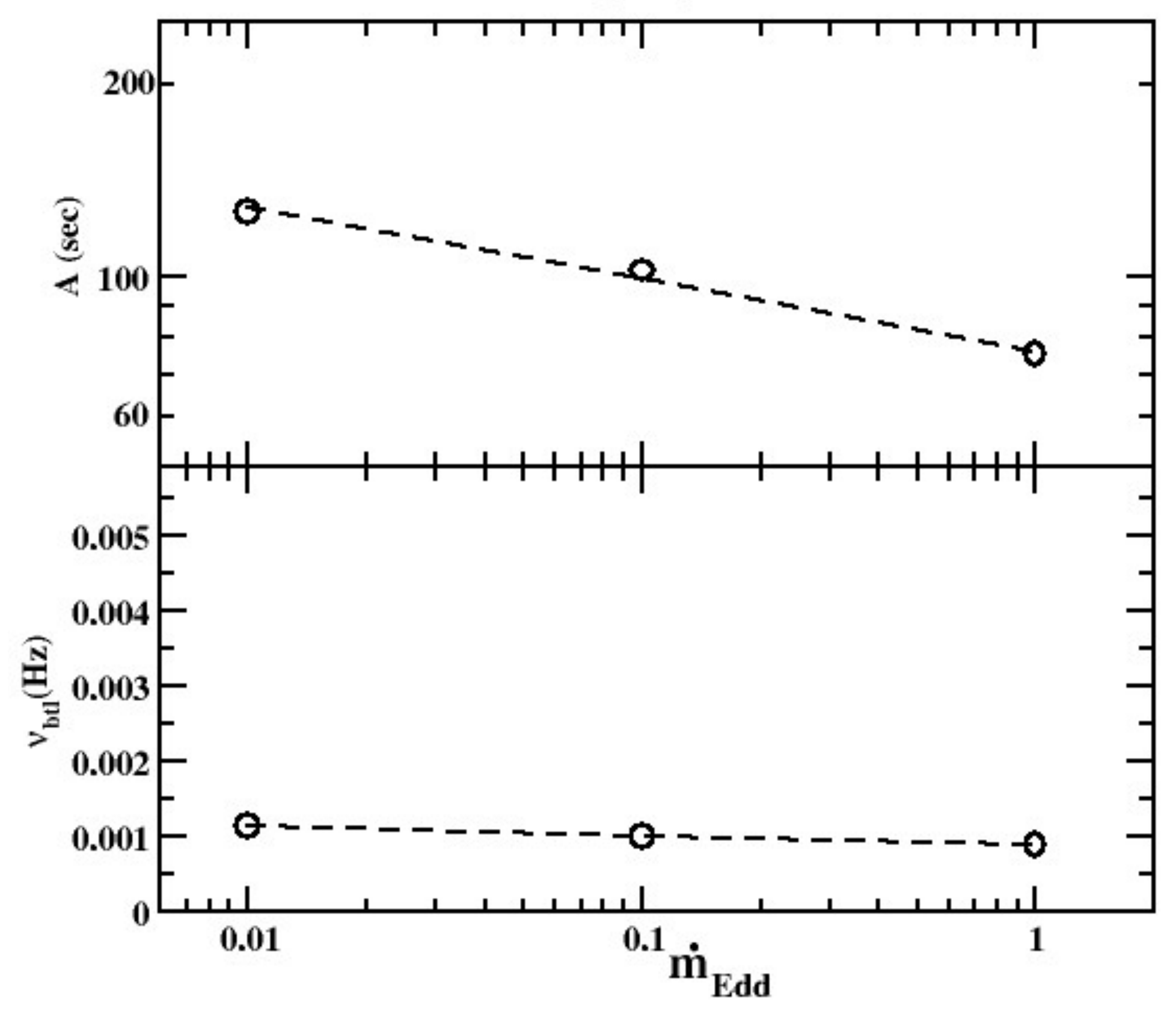}
\caption{{\it Top and lower panels:} The time-lags amplitude, $A$, and the break frequency, $\nubtl$, as a function of $\mdot$.}
\label{fig:mdotres}
\end{figure}

The number of scatterings $N_{\rm sca}(E_{\rm i})$ depends on the mean seed photon energy, $(\bar{E}_{\rm seed})$. As $\bar{E}_{\rm seed}$ increases, $\bar{N}_{\rm sca}(E_{\rm i})$ decreases. For a fixed BH mass and height, $\bar{E}_{\rm seed}$ depends on the disc accretion rate. To investigate how the time-lags depend on the accretion rate, we also compute the time-lags for $\dot{m}_{\rm Edd}=0.01$ and 1, when keeping the other coronal parameters fixed to the fiducial values listed in Table~\ref{tab:param}. \codename{} computes the energy of all photons emitted from the accretion disc when entering the corona surface. We can therefore measure their mean energy, which is  0.036, 0.065 and 0.12 keV, for $\dot{m}_{\rm Edd}=0.01,0.1$ and 1,  respectively. The resulting time-lags for the different $\bar{E}_{\rm seed}$ are plotted in Fig.~\ref{fig:mdotres}. The dependence of the time-lags ($\nubtl$ in particular) on the soft photons energy is rather weak. In general, the time-lags amplitude decreases with increasing $\dot{m}_{\rm Edd}$ (and $\bar{E}_{\rm seed}$) and the break frequency decreases with increasing $\mdot$.

The explanation for these results is straightforward. $\Delta \bar{N}_{\rm sca}(E_{\rm i}, \bar{E}_{\rm sca})$ (hence, the time-lags as well; see eq.~\ref{eq:tlagsamp}) decreases as $\bar{E}_{\rm seed}$ increases. Since less scatterings are necessary to reach higher energy, the mean light travel path between successive scatterings, $\bar{\lambda}$, is slightly larger than before. So, $\bar{\lambda}/c$ increases, and $\nubtl$ decreases with increasing $\mdot$.

The time-lags and $\nubtl$ in Fig.~\ref{fig:mdotres} are plotted as a function of the accretion rate. As we explained, the time-lags properties depend on $\bar{E}_{\rm seed}$, which in turn is set by the accretion rate. The relation between $\bar{E}_{\rm seed}$ and $\mdot$ is not unique, as it depends on the assumed colour correction term. We have used the accretion rate in all equations we give below for the time-lags and the cross-spectra, because this is the physical parameter that characterizes the accretion disc emission (for a given BH mass). We plot the accretion rate versus the mean soft photons energy that we computed above, and find that

\begin{equation}
    \mdot=5600 \bar{E}_{\rm seed}^4.
\label{eq:mdoteseed}
\end{equation}

\noindent This is the relation between $\bar{E}_{\rm seed}$ and $\mdot$ in the case of a $10^7$ M$_{\odot}$ BH, when the colour correction factor  is $f_{\rm col}=2.4$. Eq.~\ref{eq:mdoteseed} can be used to replace $\mdot$ with $\bar{E}_{\rm seed}$ in all equations we present in the following sections. This could be useful if a different colour correction term is assumed for the disc emission. If $\bar{E}_{\rm seed}$ is computed in this case, then the equations as a function of $\bar{E}_{\rm seed}$ should be used.

We note that $\bar{E}_{\rm seed}$ should also depend on the BH mass, because the Novikov-Thorne temperature profile depends both on $\mdot$ and $\mbh$ through the combination $\mdot/\mbh$ (as long as the disc is assumed to emit as black body). Hence, for a black hole with mass $\mbh$ and mass accretion rate $\mdot$, the disc temperature (as a function of distance, measured in $\Rg$) is identical with the temperature in the case of a $10^7 M_\odot$ black hole and a mass accretion rate of $\mdot / M_7$ (where $M_7\equiv \mbh / 10^7 M_\odot$). For that reason, we use the term of ($\mdot / M_7$), instead of just $\mdot$ in the equations for the time-lags parameters which we present in the following sections. In this way, the disc temperature profile is always renormalized to the temperature profile of a disc around a $10^7$ M$_{\odot}$ black hole, with the same accretion rate (in Eddington units).  Then eq.~\ref{eq:mdoteseed} could be used if a different colour correction term is assumed.

We note that $\bar{E}_{\rm seed}$ also depends on the height and radius of the corona, for a given BH mass and accretion rate. The equations we present below include the dependence of the time-lags parameters on the corona height and radius, and the respective terms account for all the effects that depend on $R_c$ and $h$, including $\bar{E}_{\rm seed}$.

\subsection{The time-lags dependence on M$_{\rm BH}$.} 
\label{sec:bhmass}

We expect the Comptonization time-lags (both the amplitude and $\nubtl$) to depend on M$_{\rm BH}$, not only because of the dependence of the disc temperature profile on $\mbh$, but also due to the dependence of $\Rg$, which is the unit length we adopt, on the BH mass. The reason for the dependence of $A$ and $\nubtl$ on $\mbh$ is that the mean photon travel time between subsequent scatterings, $\bar{\lambda}/c$, is proportional to the corona radius. Hence, $\bar{\lambda}/c$ will increase with increasing BH mass, as long as the corona radius is the same, in $\Rg$. Both the time-lags amplitude and $\nubtl$ depend on $\bar{\lambda}/c$; $A$ is proportional to $\bar{\lambda}/c$ (i.e.\ eq.~\ref{eq:tlagsamp}), while $\nubtl$ is inversely proportional to $\bar{\lambda}/c$ (see discussion in \S \ref{sec:nubreakres}). Consequently, we expect $A$ to increase proportionally with $\mbh$ and $\nubtl$ inversely proportionally with $\mbh$.


\subsection{Equations for time-lags}
\label{sec:timelagseqs}

Using the results from the best-fit models to the plots of the time-lags amplitude and break frequency versus the various physical parameters we presented in the previous sections (as indicated by the dashed-lines plotted in Figs.~\ref{fig:ampres}, \ref{fig:nubreakres}, \ref{fig:heightres} and \ref{fig:mdotres}), we reach the following equations: 

\begin{equation}
\begin{split}
    A(E_i, E_{\rm ref})=&   M_7 \left[\frac{\dot{m}_{\rm Edd} / M_7}{0.1}\right]^{-0.1} 
    \left(2.7+104\frac{\Rc}{10~\Rg}\right) \left[ \frac {\log (E_i/E_{\rm ref})}{0.23} \right]\\
   &  \cdot \left(\frac{\Te}{100\,{\rm keV}}\right)^{-0.95} \cdot 3.70 (e^{-\frac{h}{1.9~\Rg}}+0.27) ({\rm sec}), 
   \label{eq:tlampall}
\end{split}
\end{equation}
\noindent and, 
\begin{equation}
\begin{split}
   \nubtl(E_i, E_{\rm ref})=& M_7^{-1}\left[\frac{\dot{m}_{\rm Edd} / M_7}{0.1}\right]^{-0.056} \\
    & \cdot N(E_i,E_{\rm ref}) \left(\frac{\Rc}{10~\Rg}\right)^{-\beta(E_i,E_{\rm ref})} \\
    & \cdot [0.7+0.3 (\taut/0.5)]\cdot  0.68(h/\Rg)^{0.13} ({\rm Hz}), \\
\end{split}
    \label{eq:nubreakall}
\end{equation}
\noindent where, 
\begin{eqnarray}
\begin{split}
    N(E_i,E_{\rm ref})=&[9.7+3\times \log(E_i/E_{\rm ref})]\times 10^{-4}, \\
    \beta(E_i,E_{\rm ref})=&0.94-0.084\times \log(E_i/E_{\rm ref}).\\
\end{split}
\label{eq:nubres}
\end{eqnarray}

Eqs.~\ref{eq:tlampall} and \ref{eq:nubreakall} give the time-lags due to thermal Comptonization only, in the case of a stationary X-ray corona that is located at height, $h$, above the BH, and the incoming soft photons are variable. $\Rc$ and $h$ are measured in $\Rg$, the temperature is in keV, M$_7$ is the BH mass in units of 10$^7$ M$_{\odot}$, and $\mdot$ is the disc accretion rate in Eddington units. In this case, $A(E_i, E_{\rm ref})$ and $\nubtl(E_i, E_{\rm ref})$ are given in seconds and Hz, respectively. 

The soft photon flux is variable in AGNs. For example, the last few years, quite a few Seyferts have been monitored for long periods, simultaneously in the optical/UV and in the X--rays. In all cases, the optical/UV flux is variable \citep[see for example Fig.~1 in][]{edelson_first_2019}. However, in all cases, the X--ray variability amplitude is significantly larger than the variability amplitude in the optical/UV bands. This implies that the X-ray source should be dynamic, i.e. its properties should vary with time. We consider the energy spectra produced by \codename{} for all the physical parameters listed in Table \ref{tab:param}, and compute the 2--10 keV band luminosity that an observer will detect by integrating the energy spectra. Then, we plot the ratio $\lambda_{2-10}=L_{2-10 {\rm keV}}/L_{\rm Edd}$ as a function of the corona parameters, and we find that

\begin{equation}
\label{eq:lambdax}
\begin{split}
\lambda_{2-10} & =10^{-3} \left (\frac{R_c}{10~\Rg} \right )^2  \left (\frac{kT_e}{100~\rm keV} \right )^3 \left ( \frac{h}{20~\Rg} \right)^{-1.4} 
  \left( \frac{\taut}{0.5} \right )^{1.7}  \left ( \frac{\dot{m}_{\rm Edd}}{0.1} \right ),
\end{split}
\end{equation}

\noindent 
The above equation gives the ratio of the $2-10~\rm keV$ luminosity over the Eddington luminosity of an X-ray corona of radius $R_c$, temperature and optical depth of $kT_e$ and $\taut$, which is located at height $h$ above the BH, and is illuminated by an accretion disc, with an accretion rate of $\mdot$. Our results show that the X-ray luminosity increases with the area, the temperature and the optical depth of the corona, as expected. It also increases with increasing accretion rate,
and it decreases with increasing height. This is because the number of disc photons which enter the corona decreases the further away the corona is.

Therefore, in addition to the variable soft photons input, the observed X--ray variations could be due to variations of the corona radius, temperature, height, and/or optical depth. Variations of any of these parameters can be modelled as a sequence of  quasi-stationary coronae, which exists for a certain life-time (representative of the characteristic time-scale over which the physical parameters vary). In this case, the X-ray emission will be the result of many ``shots", each one representing the emission from a single, quasi-stationary corona. These shots/coronae could appear randomly, with a certain average rate and life-time (duration), which may depend on their properties. For example, a larger or hotter corona may live longer/shorter, it may be less common etc. It is difficult to model the Comptonization time-lags  without a specific model that describes the variability of the corona parameters. We provide below results regarding the expected time-lags when the corona is intrinsically variable, under some simple and rather general assumptions.

\subsection{A corona with a finite duration}

The light light curves plotted in Fig.\,\ref{fig:on_lc} show the corona response when soft photons cross its boundaries at a certain time, say $t=0$. Let us denote with $f_i(t)$ the corona response in the energy band $E_i$ in this case (for example, $f_1(t)$ would be the corona response in the energy band 1--2 keV, indicated by the red line in Fig.\,\ref{fig:on_lc}). Let us also assume that the corona appears instantaneously at time $t=0$, and lasts until $t=T$, when heating stops, and the corona disappears very quickly. In this case, the corona flux in the band $E_i$ will be 
\begin{equation}
    F_i(t) = \int_{-\infty}^{+\infty} g(t') f_i(t-t')dt',
\label{eq:shotlc}
\end{equation}

\noindent where $g(t)=1$, for $0\le t \le T$, while $g(t)=0$ otherwise. The cross-covariance between the corona output in two different energy bands in this case is: 
\begin{equation}
\begin{split}
    {\rm cov}_{F_iF_j}(k)\equiv & {\rm cov}[F_i(t)F_j(t+k)]\\  
    = & \int_{-\infty}^{+\infty}F_i(t)F_j(t+k)dt\\
    = & \int_{-\infty}^{+\infty} [ \int_{-\infty}^{+\infty}g(t')f_i(t-t')]dt' \\
    & \int_{-\infty}^{+\infty} g(t'')f_j(t+k-t'')dt''] dt   \\
    = & \int_{-\infty}^{+\infty}\int_{-\infty}^{+\infty} g(t')g(t'') \\
    & [\int_{-\infty}^{+\infty}f_i(t-t')f_j(t+k-t'')dt]dt'dt''  \\
   = &\int_{-\infty}^{+\infty}\int_{-\infty}^{+\infty} g(t')g(t'') {\rm cov}_{f_if_j}(k+t'-t'') dt' dt'', \\
\end{split}
   \label{eq:covshot}
\end{equation}

\noindent where cov$_{f_if_j}(\tau)$ is the cross-covariance of the corona instantaneous responses (i.e. like the ones plotted in  Fig.\,\ref{fig:on_lc}).The cross spectrum between the output in the two bands will be given by
\begin{equation}
\begin{split}
    h_{F_i,F_j}(\omega)=& \frac{1}{2\pi}\int_{-\infty}^{+\infty}{\rm cov}_{F_iF_j}(k) e^{-i\omega k}dk \\
                       =& \frac{1}{2\pi} \int_{-\infty}^{+\infty}\int_{-\infty}^{+\infty}\int_{-\infty}^{+\infty}
                       g(t')g(t'') {\rm cov}_{f_if_j}(k+t'-t'')  \\
                       & e^{-i\omega (k+t'-t'')} 
                       e^{i\omega t'} e^{-i\omega t''} dt' dt'' dk \\
                       =& \frac{1}{2\pi} \int_{-\infty}^{+\infty} g(t')e^{i\omega t'}dt' \int_{-\infty}^{+\infty} g(t'')e^{-i\omega t''}dt''  \\
                       & \int_{-\infty}^{+\infty} {\rm cov}_{f_if_j}(k+t'-t'')e^{-i\omega (k+t'-t'')} dk \\
                       = & \frac{1}{2\pi} T^2{\rm sinc}^2(T\omega/2) h_{f_i f_j}(\omega), \\
\end{split}
   \label{eq:shotcs}
\end{equation}

\noindent where $h_{f_if_j}(\omega)$ is the cross spectrum of the corona instantaneous responses, and sinc$^2(x)=\sin^2(\pi x)/(\pi x)^2$. The equation above implies that $arg[h_{F_i,F_j}(\omega)]=arg[h_{f_i f_j}(\omega)]$, hence the time-lags between the instantaneous corona responses in two energy bands (as defined by eqs.\, \ref{eq:tlampall} and \ref{eq:nubreakall}), are still valid in the case of a corona with a finite lifetime. However, the sinc$^2(T\omega/2)$ term indicates that the amplitude of the cross-spectrum virtually decreases to zero at frequencies higher than $\sim 1/T$. If then $T$ is quite large so that $1/T<\nubtl$, the time-lags will decrease at frequencies lower than $\nubtl$. 

We note that the case of a corona which appears instantaneously at some time above the BH is not exactly equivalent to the case we studied in the previous sections. In this case, the corona is formed at a region where soft photons already exist. We show in the Appendix that the expected time-lags are very similar both in the case when the soft photons arrive, simultaneously, on the corona surface (the case we studied in the precious sections) and in the case when soft photons are distributed throughout the corona.

\subsection{Multiple X--ray coronae}
\label{sec:multi_model}

It is quite possible that X--rays from AGNs are emitted from a number of ``active" regions, i.e. X--ray emitting regions, with a distribution of parameters (radius, temperature and/or optical depth, duration, height), which may be formed at different heights above the central BH and/or the inner disc, and may appear with a rate that depends on their parameters. We present below a simplified model that describes such a scenario, within the context of a ``shot-noise" model. 

Let us assume that the X--ray flux in two energy bands, $E_1$ and $E_2$, is the sum of the flux emitted by many active regions, as follows
\begin{equation}
F_{{\rm X},E_1}(t)=\sum_{i=1}^{N}\int_{-\infty}^{t}F_{1,i}(t-\tau)dn_i(\tau), 
\label{eq:shotlci}
\end{equation}
\noindent and,
\begin{equation}
F_{{\rm X},E_2}(t)=\sum_{i=1}^{N}\int_{-\infty}^{t}F_{2,i}(t-\tau)dn_i(\tau), \\
\label{eq:shotlcj}
\end{equation}

\noindent where $F_{1,i}(t)$ and $F_{2,j}(t)$ are given by eq.\,\ref{eq:shotlc}, $N$ is the maximum number of the various active regions that can be formed, each with a probability which is set by $dn_i(t)$. This is a random variable which determines the number of the active regions with specific characteristics which occur in the interval $(t,t+\delta t)$. Ignoring the possibility of more than one X--ray emitting region appearing in an infinitesimal interval, $\delta t$, $dn(t)$ takes only two values, namely 1 (with a probability of $\lambda\delta t$), and zero with probability of $(1-\lambda\delta t)$. It follows that, $E[dn(t)]=\lambda\delta t$, and var$[dn(t)]=\lambda\delta t$. We will also assume that cov$[dn(t), dn(t')]=0$, for $t\ne t'$. This is an important assumption, which corresponds to the case of a random shot--noise model. This assumption may not be valid for AGNs. It is possible that a correlation may exist between the times the various active regions appear. For example, after a luminous/long duration source, shorter/fainter sources may be more probable etc. In the absence of such a detailed model, we adopt this simpler assumption. In any case, at this moment, we wish to understand how a basic variability model can affect the intrinsic Comptonization time-lags. 

Assuming that $F_{\rm X}(t)=0$, $t<0$, at all bands, the cross-covariance function between the light curves in the $E_i$ and $E_j$ bands will be given by 
\begin{equation}
    \begin{split}
{\rm cov}_{F_{{\rm X}, E_1}, F_{{\rm X}, E_2}}(k)= & E[\{\sum_i \int_{-\infty}^{+\infty}F_{1,i}(t-\tau)dn_i(\tau)\} \times \\
     & \{\sum_j \int_{-\infty}^{+\infty}F_{2,j}(t+k-\tau')dn_j(\tau')\} ] \\
     = & \sum_i  \sum_j  \int_{-\infty}^{+\infty} \int_{-\infty}^{+\infty} F_{1,i}(t-\tau)  F_{2,j}(t+k-\tau') \\
& \times E[dn_i(\tau)dn_j(\tau')] \\
= & \sum_i \int_{-\infty}^{+\infty} F_{1,i}(t')  F_{2,i}(t'+k) \lambda_i dt'.\\
= &   \sum_i \lambda_i{\rm cov}_{F_{1,i}F_{2,i}}(k). \\
    \end{split}
    \label{eq:shotccvf}
\end{equation}

\noindent And the cross-spectrum will be 
\begin{equation}
    \begin{split}
    h_{F_{{\rm X}, E_1}, F_{{\rm X}, E_2}}(\omega)
    =&\frac{1}{2\pi}\int_{-\infty}^{+\infty}
    {\rm cov}_{F_{{\rm X}, E_1}, F_{{\rm X}, E_2}}(k) e^{-i\omega k} dk \\
    =&\sum_i \lambda_i \left[\frac{1}{2\pi} \int_{-\infty} ^{+\infty} {\rm cov}_{F_{1,i}F_{2,i}}(k) e^{-i\omega k} dk\right] \\
    =& \sum_i \lambda_i [ h_{F_1,F_2}(\omega) ]_i \\
    = & \frac{1}{2\pi} \sum_i \lambda_i T^2_i {\rm sinc}^2 (T_i\omega/2)[ h_{f_1 f_2}(\omega)]_i. \\
    \end{split}
    \label{eq:shotcs_multi}
 \end{equation}   

\begin{figure}
 \includegraphics[width=\columnwidth]{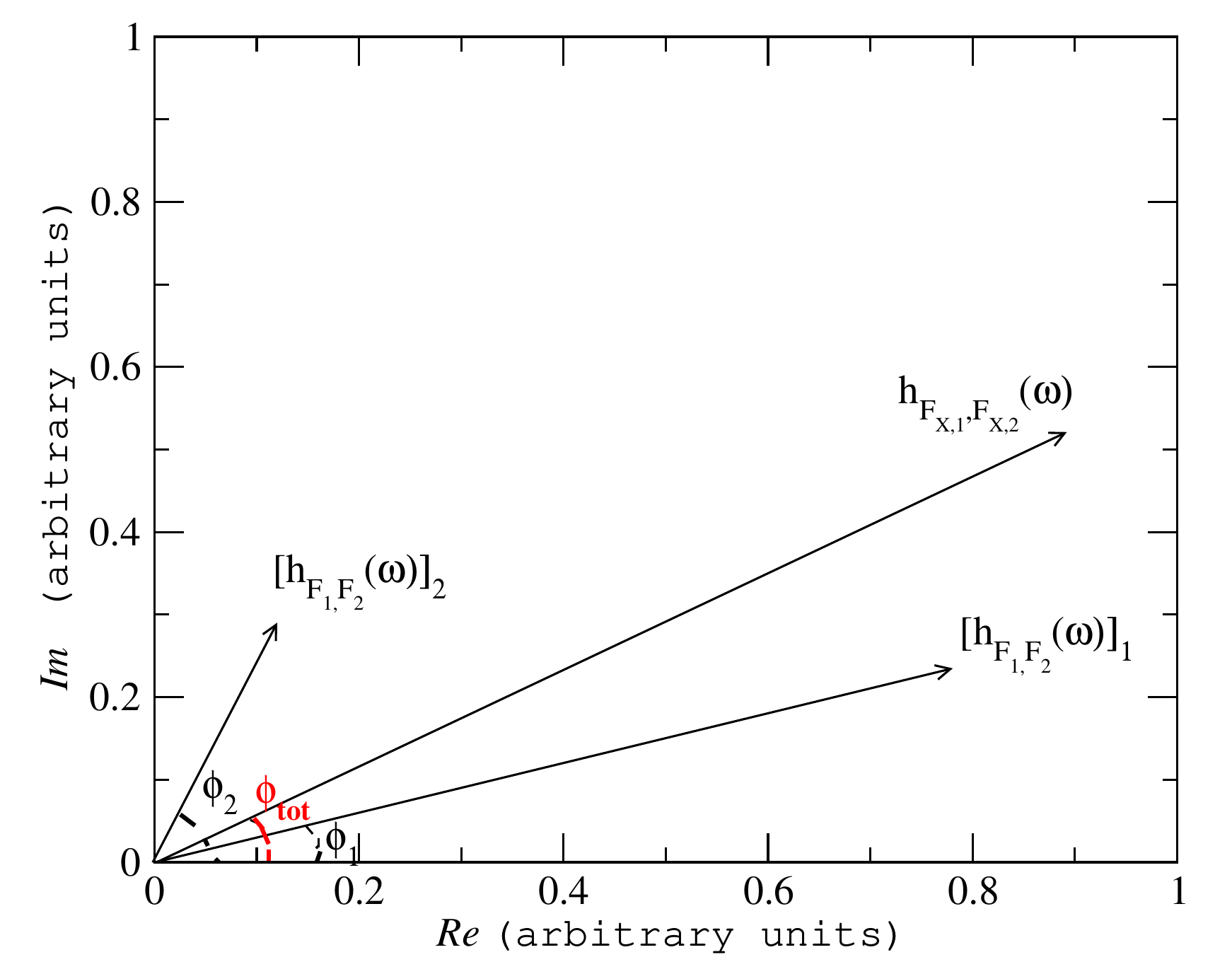}
\caption{Example of the cross-spectra of two coronae, and their sum (angles indicate the respective phase-lags). }
\label{fig:sketch}
\end{figure}

The equation above shows that the final cross-spectrum is equal to the sum of the cross-spectra of the individual X--ray coronae. Consequently, the final time-lag is not the sum of the individual time-lags; it is the cross-spectrum itself that matters in this case. An illustration of eq.\,\ref{eq:shotcs_multi} in the case of two coronae is shown in Fig.\,\ref{fig:sketch}. Vectors $ [\hat{h}_{f_1 f_2}(\omega)]_1$ and $[\hat{h}_{f_1 f_2}(\omega)]_2$ indicate the cross-spectrum of the two coronae at frequency $\omega$. Angles $\phi_1$ and $\phi_2$ are the phase lags between the variations in bands 1 and 2, for each one of the two shots. The final cross-spectrum is indicated by vector $\hat{h}_{F_{{\rm X},1}, F_{{\rm X}, 2}}(\omega)$, and the total phase lag is determined by angle $\phi_{\rm tot}$. This is closer to $\phi_1$, because the amplitude of the first cross-spectrum is larger than that of $[\hat{h}_{f_1 f_2}(\omega)]_2$. In principle, in addition to the physical parameters of the X--ray corona, the amplitude of each $[\hat{h}_{f_1 f_2}(\omega)]_i$ should also depend on how often each corona appears (i.e on $\lambda_i$) and on how long it lasts (i.e. on $T_i$).

\subsection{The cross-spectrum of the X--ray corona}

\begin{figure}
    \centering
    \includegraphics[width=\columnwidth]{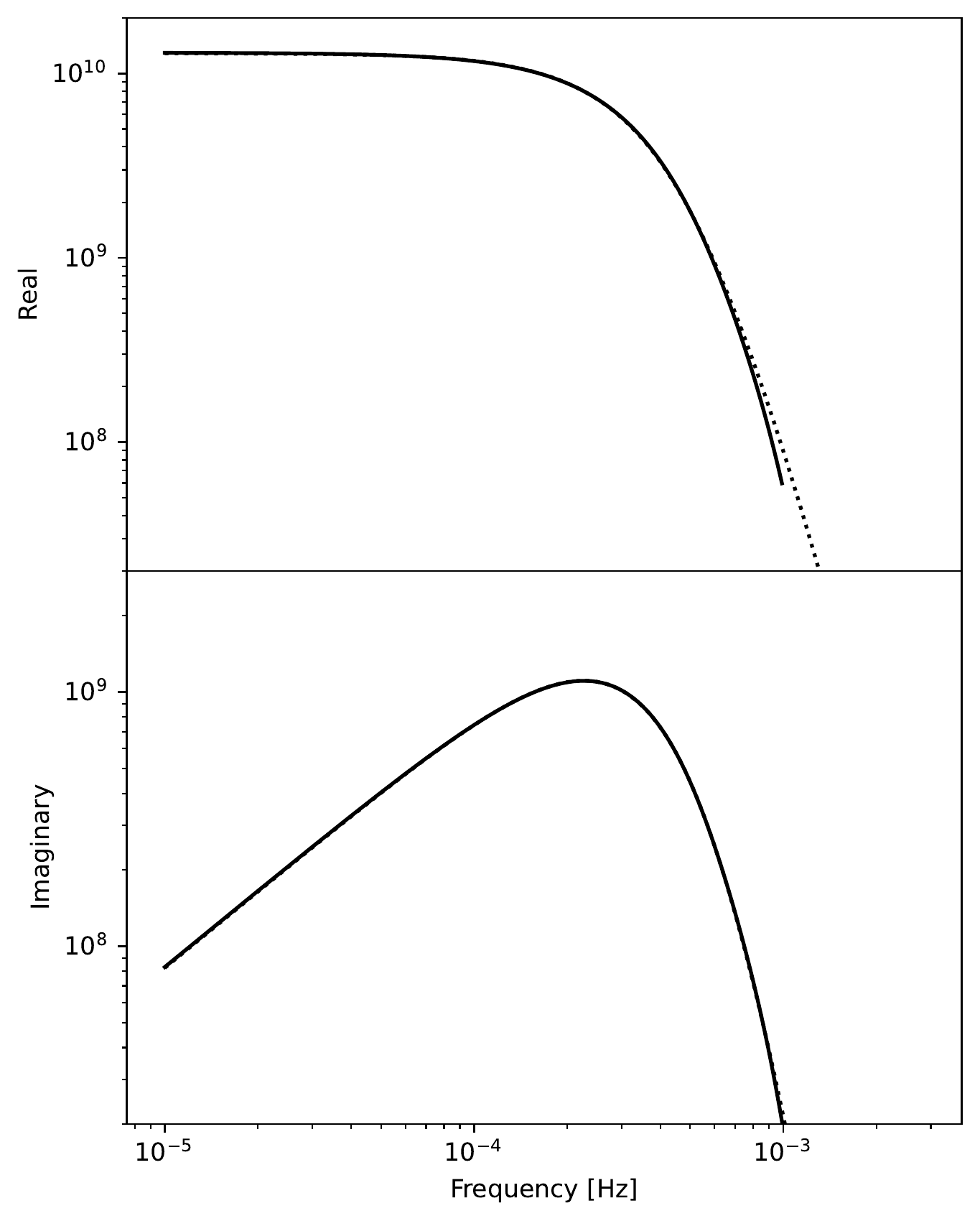}
    \caption{The best-fit model (solid lines) to the real and imaginary part of the cross-spectrum (dotted lines; top and bottom panels, respectively) in the case of a single corona with $h=20~\Rg$,$\Rc=10~\Rg$,$\taut=0.5$, and $kT_e$=100 keV. }
    \label{fig:fitrealimg}
\end{figure}

As we discussed above, to compute the time-lags of multiple X-ray coronae we need to sum cross-spectra of the coronal responses to the soft input photons, like the ones plotted in Fig.~\ref{fig:on_lc}. As an example of how these cross-spectra look like, the dotted lines in Fig.~\ref{fig:fitrealimg} show the real and imaginary parts
of the cross spectrum between the lightcurves in the 4--6 and 2--4 keV bands for the corona we studied in \S~\ref{sec:example} (i.e. $\Rc=10~\Rg$, $h=20~\Rg$, $\taut=0.5$, and $\Te=100~\rm keV$). The real and imaginary parts of the cross spectrum, $\Re(\nu)$ and $\Im(\nu)$ are constant and increase proportional to $\nu$, respectively, at frequencies below a break frequency, which is almost the same for both of them (i.e. $\nu_{0r}\approx \nu_{0i})$. Actually, it is this difference in the dependence on frequency between $\Re(\nu)$ and $\Im(\nu)$ that results in the time-lags being constant below $\nu_{\rm btl}$. At higher frequencies, both $\Re(\nu)$ and $\Im(\nu)$ decrease exponentially with increasing frequency. Based on the shape of the curves plotted in Fig.~\ref{fig:fitrealimg}, we fit them with the following models:

\begin{eqnarray}
\label{eq:modelreal}
 \Re(\nu) &=& A_{0r}\left[e^{-\left(\frac{\nu}{\nu_{0r}}\right)^{\alpha_r}} + 
         0.25 e^{-\left(\frac{\nu}{1.58\nu_{0r}}\right)^{\alpha_r}}\right],\\
\label{eq:modelim}
 \Im(\nu) &=& \nu \cdot A_{0i}\left[e^{-\left(\frac{\nu}{\nu_{0i}}\right)^{\alpha_i}} + 
         0.18 e^{-\left(\frac{\nu}{1.48\nu_{0i}}\right)^{\alpha_i}}\right].
\end{eqnarray}
The solid lines in Fig.~\ref{fig:fitrealimg} shows the best-fit lines to the real and imaginary parts of the cross-spectrum. Clearly, the models given by the equations above fit $\Re(\nu)$ and $\Im(\nu)$ well.

We proceed with computing the cross spectra of all disc-corona systems (listed in Table \ref{tab:param}) and fitting them with eqs.~\ref{eq:modelreal} and \ref{eq:modelim}. Then, we use the best-fit results to study how the $\Re(\nu)$ and $\Im(\nu)$ model parameters depend on the corona physical parameters. Our results are as follows:

\begin{equation}
\begin{split}
 A_{0r} =& 1.58\cdot 10^{10} M_7
    \left(\frac{\mdot/M_7}{0.1}\right)^{2.12}
    \left(\frac{R_c}{10~\Rg}\right)^{5.78}
    \left(\frac{kT_e}{100~\rm keV}\right)^{5.54} \\
    & \left(\frac{\tau}{0.5}\right)^{3.08}
    \left(\frac{h}{20~\Rg}\right)^{-3.88},
\end{split}
\label{eq:cross_amp_real}
\end{equation}
\begin{equation}
\begin{split}
\label{eq:nu0r}
\nu_{0r} =& 3.12\cdot 10^{-4} M_7^{-1}
    \left(\frac{\mdot/M_7}{0.1}\right)^{0.04}
    \left(\frac{R_c}{10~\Rg}\right)^{-0.94} \left(\frac{kT_e}{100~\rm keV}\right)^{0.30}\\
    &  (1-1.30e^\frac{-h}{3.08~\Rg})~\rm (Hz), 
\end{split}
\end{equation}
\begin{equation}
\begin{split}
\label{eq:ar}
\alpha_r =& 1.89
    \left(\frac{\mdot/M_7}{0.1}\right)^{-0.01}
    \left(\frac{R_c}{10~\Rg}\right)^{-0.03}
    \left(\frac{kT_e}{100~\rm keV}\right)^{-0.05}
    \left(\frac{\tau}{0.5}\right)^{0.02}\\
    & (1+2.58e^\frac{-h}{0.65~\Rg}),
\end{split}
\end{equation}
and, 
\begin{equation}
\begin{split}
 A_{0i} =& 1.12\cdot 10^{13} M_7^2 
    \left(\frac{\mdot / M_7}{0.1}\right)^{2.01}
    \left(\frac{R_c}{10~\Rg}\right)^{6.75}
    \left(\frac{kT_e}{100~\rm keV}\right)^{4.46}\\
    & \left(\frac{\tau}{0.5}\right)^{3.22}
    \left(\frac{h}{20~\Rg}\right)^{-4.24},
\end{split} 
\label{eq:cross_amp_imag}
\end{equation}
\begin{equation}
\begin{split}
\label{eq:nu0i}
\nu_{0i} = & 2.94\cdot 10^{-4} M_7^{-1} 
    \left(\frac{\mdot / M_7}{0.1}\right)^{0.02}
    \left(\frac{R_c}{10~\Rg}\right)^{-0.97}
    \left(\frac{kT_e}{100~\rm keV}\right)^{0.23} \\
    & (1-1.18e^\frac{-h}{3.36~\Rg})~\rm (Hz),\\
\end{split}
\end{equation}
\begin{equation}
\begin{split}
\label{eq:ai}
\alpha_i =& 1.85
    \left(\frac{\mdot / M_7}{0.1}\right)^{-0.01}
    \left(\frac{R_c}{10~\Rg}\right)^{-0.04}
    \left(\frac{kT_e}{100~\rm keV}\right)^{-0.01}
    \left(\frac{\tau}{0.5}\right)^{0.01} \\
    & (1+4.74e^\frac{-h}{0.63~\Rg}),
\end{split}
\end{equation}
while $\nu_{0r}$ and $\nu_{0i}$ should be multiplied by a factor of $0.91$ and $0.94$, respectively, if $\taut\geq 1$.

Eqs.~\ref{eq:modelreal} and \ref{eq:modelim}, together with eqs.\,\ref{eq:cross_amp_real}--\ref{eq:ai}, can be used to compute the real and imaginary parts of the cross-spectrum between 2--4 and 4--6 keV band light curves, due to thermal Comptonization, for a corona with parameters ($\Rc,kT_e, \taut$), located at height $h$ above the BH, with mass $M_7$, when the mass accretion rate is $\mdot$.
We provide results for the cross spectrum between the $2-4$ and $4-6$ keV bands only, as these should be the best possible choices for the study of time-lags in AGN, as they are relatively unaffected by the X-ray reflection components.

\section{Practical implications of our results}
\label{sec:implications}
As an example of how our work can be used in practise, we consider the results of \citetalias{epitropakis_x-ray_2017}, who used {\it XMM-Newton} data for 10 X-ray bright and highly variable AGNs to measure the X--ray continuum time-lags.  
According to \citetalias{epitropakis_x-ray_2017}, the time-lags between two energy bands with mean energy of $E_1$ and $E_2$ are given by:
\begin{equation}
\label{eq:ep17}
\tau_{\rm obs}(\nu)=10^{\alpha}(\lambda_{2-10})^{\beta}
\log \left (\frac{\rm E_2}{\rm E_1}\right)
\left(\frac{\nu}{10^{-4} \rm Hz}\right)^{-1}\ (\rm sec),
\end{equation}

\noindent where  $\alpha=3.42 \pm 0.13$, and $\beta=0.55 \pm 0.07$. The thick solid and dotted lines in Figs.~\ref{fig:obs1} -- \ref{fig:multi_shot} show the \citetalias{epitropakis_x-ray_2017} best-fit and the upper 1 and $3\sigma$ results, respectively, while the two vertical lines indicate the frequency range of $10^{-4}- 10^{-3}$ Hz. This is (approximately) the frequency range where \citetalias{epitropakis_x-ray_2017} estimated the X--ray continuum time-lags. Strictly speaking, this is also the frequency range where the EP17 results are also valid. We also assumed that $\lambda_X=0.015$, which is the median luminosity of AGNs in the \citetalias{epitropakis_x-ray_2017} sample. We present below a comparison between the model thermal Comptonization time-lags (both for a single and multiple coronae) and the \citetalias{epitropakis_x-ray_2017} results.

\subsection{A single corona}
\label{sec:single_corona}

\begin{figure}
\includegraphics[width=\columnwidth]{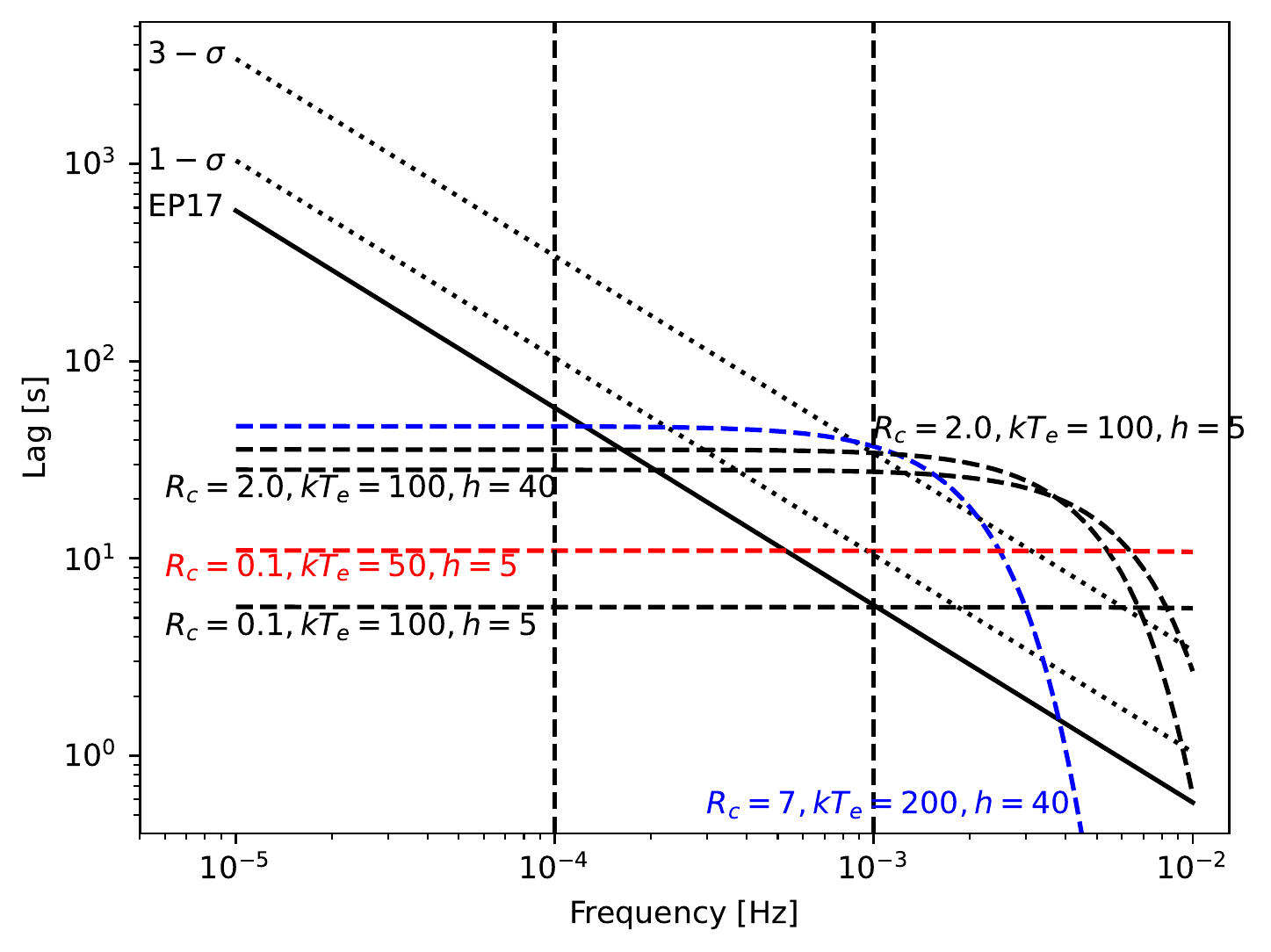}
\caption{Observed vs model Comptonization time-lags. Solid and dotted black lines show the \citetalias{epitropakis_x-ray_2017} time-lags for $\lambda_X=0.015$, and the 1 and 3$\sigma$ upper limits, respectively. The dashed lines show the Comptonization time-lags for various coronal parameters, as indicated in the plot. Model lines for  $\Te=50, 100,$ and 200 keV are plotted in red, black, and blue colors, respectively. The vertical lines in this figure, as well as in Figs.~\ref{fig:multi_uniform}, \ref{fig:multi_dist}, and \ref{fig:multi_shot}, indicate the frequency range over which \citetalias{epitropakis_x-ray_2017} measured time-lags. 
 \label{fig:obs1}}
\end{figure}

\begin{figure}
 \includegraphics[width=\columnwidth]{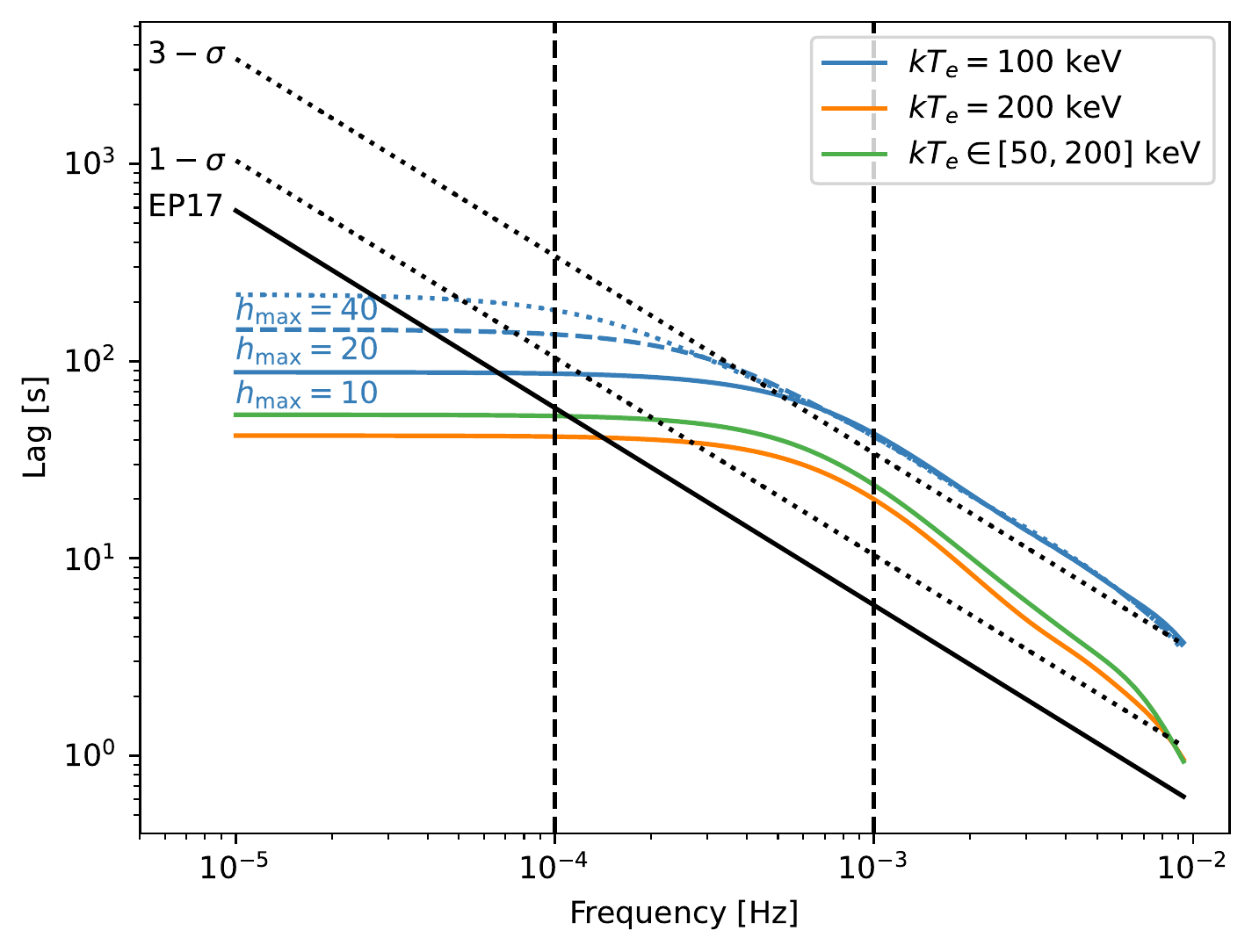}
 \caption{The time-lags of multiple coronae (which have the same duration). Both the height and radius of the coronae are uniformly distributed. The results for $h_{\rm max}=10~\Rg$ are plotted in solid curves. Time-lags when  $\Te=100~\rm keV$, $\Te=200~\rm keV$, or $\Te$ is uniformly distributed between $50$ and $200~\rm keV$, are plotted in blue, orange, and green colors, respectively. We also plot the time-lags when $\Te=100~\rm keV$, $h_{\rm max}=20$ and $40~\Rg$, in blue dashed and dotted lines, respectively.
 \label{fig:multi_uniform}}
\end{figure}

We start with the time-lags of a single, stationary corona. In Fig.~\ref{fig:obs1} we compare the model time-lags of a few single coronae (computed with eqs.~\ref{eq:tlampall}--\ref{eq:nubreakall}) with the \citetalias{epitropakis_x-ray_2017} results. 
It is apparent that a single corona cannot reproduce the observations. This was already obvious by looking at eq.~\ref{eq:model}, which shows that the thermal Comptonization time-lags of a single, stationary corona remain constant below $\nubtl$ and then decay exponentially above $\nubtl$. This shape is different than the powerlaw profile of the observed time-lags. 

Although a single corona cannot explain the observed time-lags, the predicted model time-lags  should not be larger than the observed time-lags at frequencies $10^{-4}-10^{-3}$ Hz, if thermal Comptonization is the X-ray emission mechanism in AGNs, since other physical processes only increase hard lags, not decrease them. To demonstrate this issue, we use eqs.~\ref{eq:model} and \ref{eq:tlampall}--\ref{eq:nubres} to compute the Comptonization time-lags between 4--6 and 2--4 keV (i.e $E_1=5$ and $E_2=3$ keV), for a corona with $h=5~\Rg$, $\Rc=0.1~\Rg$, and $\Te=50$ keV ($\taut=1$, $\mdot=0.1$, and $\mbh=10^7~\rm M_\odot$ for all model time-lags in this figure). The model (shown with the red-dashed line) crosses the observed time-lags at $\sim 5-6\times 10^{-4}$ Hz. This implies that such a corona cannot be responsible for the X-ray emission from AGNs. If that was the case, the observed time-lags would not have a power-law shape up to $10^{-3}$ Hz. The time-lags are smaller and $\nubtl$ is higher for a corona with the same size and height, but which is hotter ($\Te=100$ keV). The time-lags in this case (bottom black dashed line in Fig.~\ref{fig:obs1}) could be in agreement with the observations but that depends on whether the intrinsic time-lags in AGNs continue with a power-law form at frequencies higher than $10^{-3}$ Hz or not. This is not clear from the observations, but we suspect this is rather unlikely, also if we consider the observed time-lags in XRBs, which extend over two or more decades of frequency with a power-law shape. 

Fig.~\ref{fig:obs1} shows that $\Rc$ can be up to $\sim 2~\Rg$ if $\Te=100$ keV and $h=5~\Rg$. Larger radii will result in time-lags which will not be consistent even with the $3\sigma$ upper limit of the observed time-lags (the top black dashed line in Fig.~\ref{fig:obs1}). The model time-lags of a $\Rc=2~\Rg, \Te=100$ keV corona agrees slightly better with the $3\sigma$ upper limit when $h=40~\Rg$, because the model amplitude decreases (the middle black dashed line in Fig.~\ref{fig:obs1}). A corona with a larger radius can agree better with the data, but only for the higher temperature of $\Te= 200$ keV (the blue dashed line in the same figure). However, even in this case, the agreement is at the $3\sigma$ level. We will need a larger height, and a hotter corona, to accommodate a larger radius corona, but again, this will be at the $3\sigma$ level, minimum. 

\subsection{Multiple coronae}
We proceed to multiple coronae, making use of eq.~\ref{eq:shotcs_multi}. With no prior knowledge of the distribution of the size and height of the corona, we start with the simplest hypothesis that they are uniformly distributed. For each corona, we first sample its height, $h$, from a uniform distribution between $3~\Rg$ and a maximum value, $h_{\rm max}$. Then we sample its radius from a uniform distribution between $1~\Rg$ and $(h_{\rm max} - 1)$. For the other parameters we assume $\taut=0.5$, $\mdot=0.1$, and $\mbh=10^7~\rm M_\odot$. We also assume the same probability, $\lambda$, and life-time for the coronae. We choose a life-time of $T=100$ s, arbitrarily, but this choice does not affect our results. If the life-time is the same for all coronae, eq.~\ref{eq:shotcs_multi} shows that the phase lag at frequency $\omega$ will be $\propto {\rm tan}^{-1}\{\sum_i \Im_i[h_{f_1f_2}(\omega)]/\sum_i \Re_i[h_{f_1f_2}(\omega)]\}$, i.e. it will be independent of the corona life-time.

Fig.~\ref{fig:multi_uniform} shows the results for various $\Te$ and $h_{\rm max}$. The amplitude of the time-lags decreases with increasing $\Te$, as before. Interestingly, the time-lags in this case do not decrease exponentially with frequency at high frequencies, but follow a power-law like profile (with a slope of $\sim -1$). To understand this, we compare (for example)  the $R_c=2~\Rg, h=40, kT_e=100~\rm keV$ curve in Fig.~\ref{fig:obs1}, with the blue dotted line in Fig.~\ref{fig:multi_uniform}. At the highest frequency (10$^{-2}$ Hz), the time-lags are comparable in both curves. The time-lags become constant at frequencies lower than $\sim 6-7\times 10^{-3}~\rm Hz$ for a single corona, but they follow a power-law shape down to $\sim 3\times 10^{-4}~\rm Hz$ in Fig.~\ref{fig:multi_uniform}. This is because the real and imaginary part at frequency $\nu$ in the latter case is mainly determined by coronae with $\nu_{0r}\approx \nu_{0i}\approx \nu$, as we sexplain below.

The real and imaginary part of all the coronae with $\nu_{0r}\approx \nu_{ir}\ll \nu$ is very small at frequency $\nu$, due to the very steep decrease of $\Re(\nu)$ and $\Im(\nu)$ at frequencies higher than $\nu_{0r}$ and $\nu_{0i}$, respectively (see Fig.~\ref{fig:fitrealimg}). As a result, these coronae do not contribute significantly to the sum of $\Re_{i}(\nu)$ and $\Im_{i}(\nu)$. On the other hand, for all the coronae with $ \nu_{0r} \approx \nu_{ir} \gg \nu$, the imaginary part at frequency $\nu$ will be very small (again because $\Im(\nu)$ decreases steeply at frequencies lower than $\nu_{0i}$ - see bottom panel in Fig.~\ref{fig:fitrealimg}). However, $\Re(\nu)$ is constant at frequencies lower than $\nu_{0r}$. Since $\nu_{0r}$ 
depends mainly on $R_c$ (eq.~\ref{eq:nu0r}), 
the radius of the coronae with $\nu_{0r}$ significantly higher than $\nu$ will be smaller than the radius of the corona with $\nu_{0r} \approx \nu$. Given the steep dependence of the amplitude of $\Re(\nu)$ on $R_c$ (eq.~\ref{eq:cross_amp_real}), the contribution of the real part of these coronae to the sum of $\Re_i(\nu)$ will also be very small. Therefore we expect that
\begin{equation}
\begin{split}
{\tau}(\nu) = \frac{1}{2\pi\nu} {\rm tan}^{-1}\left[\frac{\sum_i \Im_i(\nu)}{\sum_i \Re_i(\nu)}\right]& \approx \frac{1}{2\pi\nu} {\rm tan}^{-1}\left[\frac{\Im_{\nu_{0i}=\nu}(\nu)}{\Re_{\nu_{0r}=\nu}(\nu)}\right] \\
& \approx \frac{1}{2\pi \nu} {\rm tan}^{-1}\left[\frac{\nu A_{0i}}{A_{0r}}\right],
\end{split}
\label{eq:taufinal}
\end{equation}

\noindent where $A_{0r}$ and $ A_{0i}$ are the amplitude of the real and imaginary parts of the cross-spectrum of the corona with $\nu_{0r}\approx \nu_{0i}\approx \nu$. Using eqs.~\ref{eq:cross_amp_real}, \ref{eq:nu0r}, and \ref{eq:cross_amp_imag}, we can see that $\nu A_{0i}/A_{0r} \approx \nu_{0r} A_{0i} / A_{0r} \propto (kT_e h)^{-1}$. Since neither $\nu_{0r}$ nor $\nu_{0i}$ depend significantly on $kT_e$ and $h$, it turns out that ${\rm tan}^{-1}(\nu A_{0i}/A_{0r}) \propto {\rm constant}$, and hence eq.~\ref{eq:taufinal} shows that $\tau(\nu) \propto \nu^{-1}$. The power-law like time-lags extend to $\nu\approx \nu_{0r}$ of the largest radius corona, since $\nu_{0r}$ (and $\nu_{0i}$) is mainly determined by $\Rc$. At even lower frequencies, the time-lags should be constant. At these frequencies,
\begin{equation}
 \tau(\nu) \propto \frac{1}{\nu}{\rm tan}^{-1} \left[\frac{\nu A_{0i}(R_{c,{\rm max}})}{A_{0r}(R_{c,{\rm max}})} \right] \approx \frac{A_{0i}(R_{c, {\rm max}})}{A_{0r}(R_{c,{\rm max}})}.
\end{equation}


We also considered the case when $\Te$ is uniformly distributed between $50$ and $200~\rm keV$ (solid green curve). The resulting time-lags are close to the case when  $\Te=200~\rm keV$ is the same for all coronae (orange curve in Fig.~\ref{fig:multi_uniform})
The reason is that, due to the steep dependence of the cross spectrum on $\Te$ (eqs.~\ref{eq:cross_amp_real} \& \ref{eq:cross_amp_imag}), the time-lags in this case are dominated by the coronae with the highest temperature. 

\begin{figure}
 \includegraphics[width=\columnwidth]{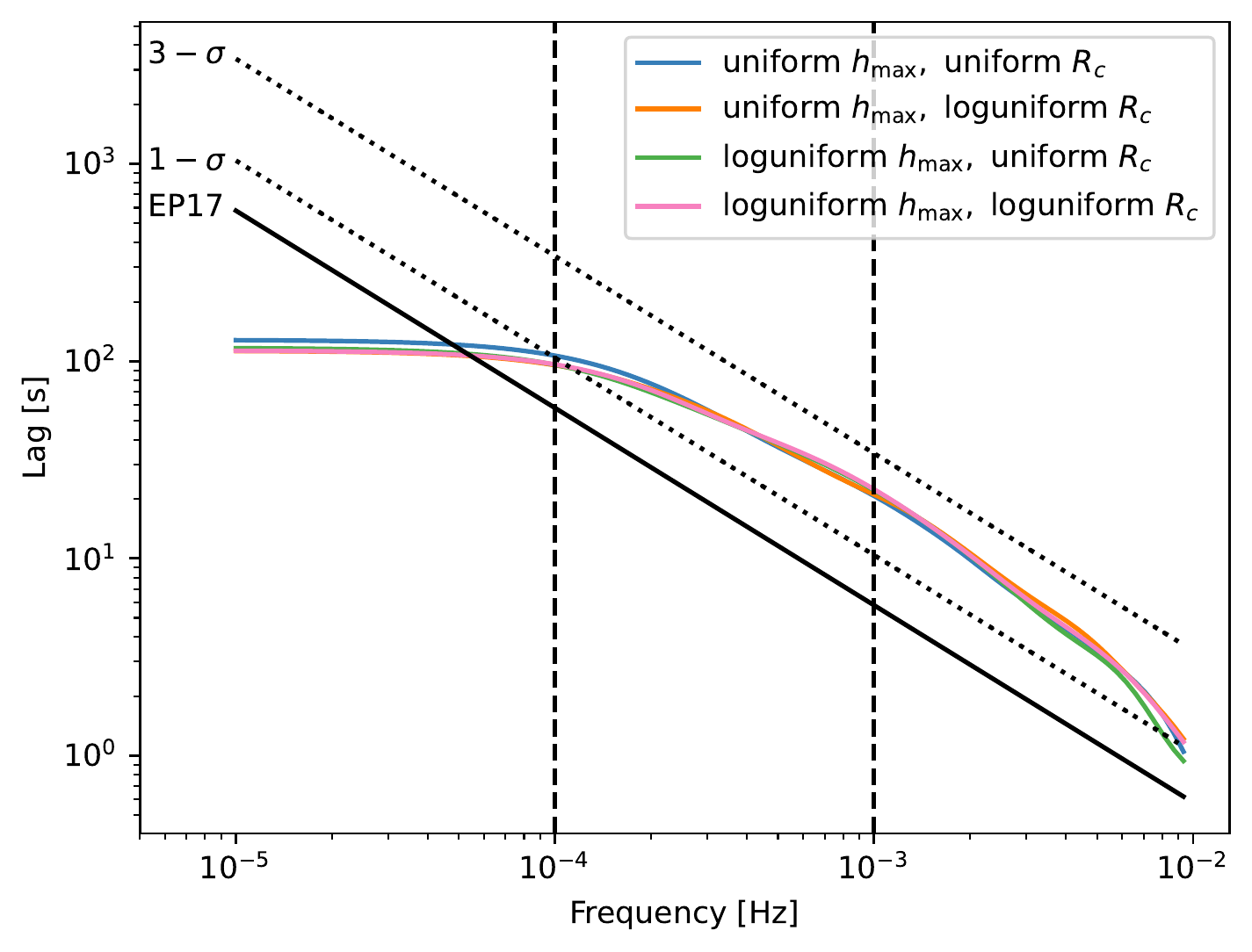}
 \caption{A comparison of the time-lags of multiple coronae with different distributions of the maximum height $h_{\rm max}$ and $\Rc$. The results for different distributions are plotted in different colors, as indicated in the plot.\label{fig:multi_dist}. The temperature follows a uniform distribution between 50--200 keV in all cases, and duration is the same for all coronae.}
\end{figure}

We go beyond a simple uniform distribution for $h_{\rm max}$ and $\Rc$ and experiment with the possibility that the distribution of the two parameters is uniform in the logarithmic scale. In Fig.~\ref{fig:multi_dist} we present the results when $h_{\rm max}$ and $\Rc$ are distributed either uniformly or loguniformly\footnote{By ``loguniform'' we mean that the logarithm of the variable is distributed uniformly.}  between 3 and $40~\Rg$, and between $1$ and $(h-1)~\Rg$, respectively. We also assume that $\Te$ follows a uniform distribution between $50$ and $200~\rm keV$. We find that the time-lags are rather insensitive to the distribution of $h$ and $\Rc$. This is not surprising though, given the steep dependence of the cross spectrum on $\Rc$.

The model time-lags in Fig.~\ref{fig:multi_dist} are closer to the observed time-lags. They can be compared 
with the model plotted with the blue dotted line in Fig.~\ref{fig:multi_uniform}. The distribution of the corona height and radius is the same in both cases. The difference is that $kT_e$ is fixed at 100 keV in Fig.~\ref{fig:multi_uniform}, while we assume a uniform distribution of $kT_e$ between 50--200 keV for the time-lags plotted in Fig.~\ref{fig:multi_dist}. As we said above, the time-lags at each frequency are mainly determined by the time-lags of coronae with $\nu_{0r}\approx \nu_{0i} \approx \nu$. But now there is a large number of coronae, each with a different temperature, for which $\nu_{0r}\approx \nu_{0i} \approx \nu$ (since $\nu_{0r}$ and $\nu_{0i}$ do not depend strongly on $kT_e$). However, $A_{0r}$ and $A_{0i}$ depend strongly on the corona temperature (eqs.~\ref{eq:cross_amp_real} and \ref{eq:cross_amp_imag}). Therefore, the time-lags at each frequency will be almost equal to the time-lags of the corona with the highest temperature (for a given radius and height). Actually, since $A_{0i}/A_{0r}\propto kT_e^{-1}$ and  $kT_e=100$ keV in Fig.~\ref{fig:multi_uniform}, while $kT_{e,{\rm max}}=200~\rm keV$ for the time-lags in Fig.~\ref{fig:multi_dist}, their amplitude should be be smaller by a factor of $\sim 0.5$ than the time-lags in Fig.~\ref{fig:multi_uniform}, as is the case.

Finally we take the life-time of the coronae into account, and we assume that duration of each corona 
is set by its physical properties. A natural choice for the life-time of an X-ray corona is the radiative cooling timescale:
\begin{equation}
\label{eq:tcool}
t_{\rm cool} \equiv \frac{E_{\rm k}}{L_{\rm cool}},
\end{equation}
where $E_{\rm k} \equiv 2 \pi R_c^3n_e kT_e$ is the thermal energy of the corona, and $L_{\rm cool}$ is the luminosity of radiative cooling. We compute $L_{\rm cool}$ by the following equation:
\begin{equation}
    L_{\rm cool} \equiv L_{\rm corona} - L_{\rm seed},
\end{equation}
where $L_{\rm corona}$ is the full-band total luminosity of the corona with contributions from both unscattered and scattered photons, and $L_{\rm seed}$ is the luminosity of the seed photons that enter the corona. We consider the energy spectra
produced by \codename{} for all the physical parameters listed in Table \ref{tab:param}, and we compute $L_{\rm corona}$, $L_{\rm seed}$ and $L_{\rm cool}$ in each case. Then, we plot the ratio $\lambda_{\rm cool}\equiv L_{\rm cool}/L_{\rm Edd}$ as a function of the corona parameters, and we find that,
\begin{equation}
\begin{split}
\lambda_{\rm cool} \equiv L_{\rm cool}/L_{\rm Edd} = & \ 5.1\cdot10^{-3} \left(\frac{\Te}{100~\rm keV}\right)^{2.3} \left(\frac{h}{20~ \Rg}\right)^{-1.7}  \\ & \left(\frac{\taut}{0.5}\right)^{2.6} \left(\frac{\dot{m}_{\rm Edd}}{0.1}\right).
\label{eq:tcooleq}
\end{split}
\end{equation}
Now we have the expression for $t_{\rm cool}$:
\begin{equation}
\begin{split}
t_{\rm cool} = & 25.7\ M_7 \left(\frac{\Te}{100~\rm keV}\right)^{-1.3} \left(\frac{h}{20~ \Rg}\right)^{1.7} \left(\frac{\taut}{0.5}\right)^{-1.6} \\
 &   \left(\frac{\dot{m}_{\rm Edd}}{0.1}\right)^{-1}\ \rm (sec).
\end{split}
\end{equation}
The equation above give the cooling time-scale of an X-ray corona of radius $R_c$, temperature and optical depth of $kT_e$ and $\taut$, which is located at
height $h$ above the BH, and is illuminated by an accretion disc, with an accretion rate of $\mdot.$
 
\begin{figure}
 \includegraphics[width=\columnwidth]{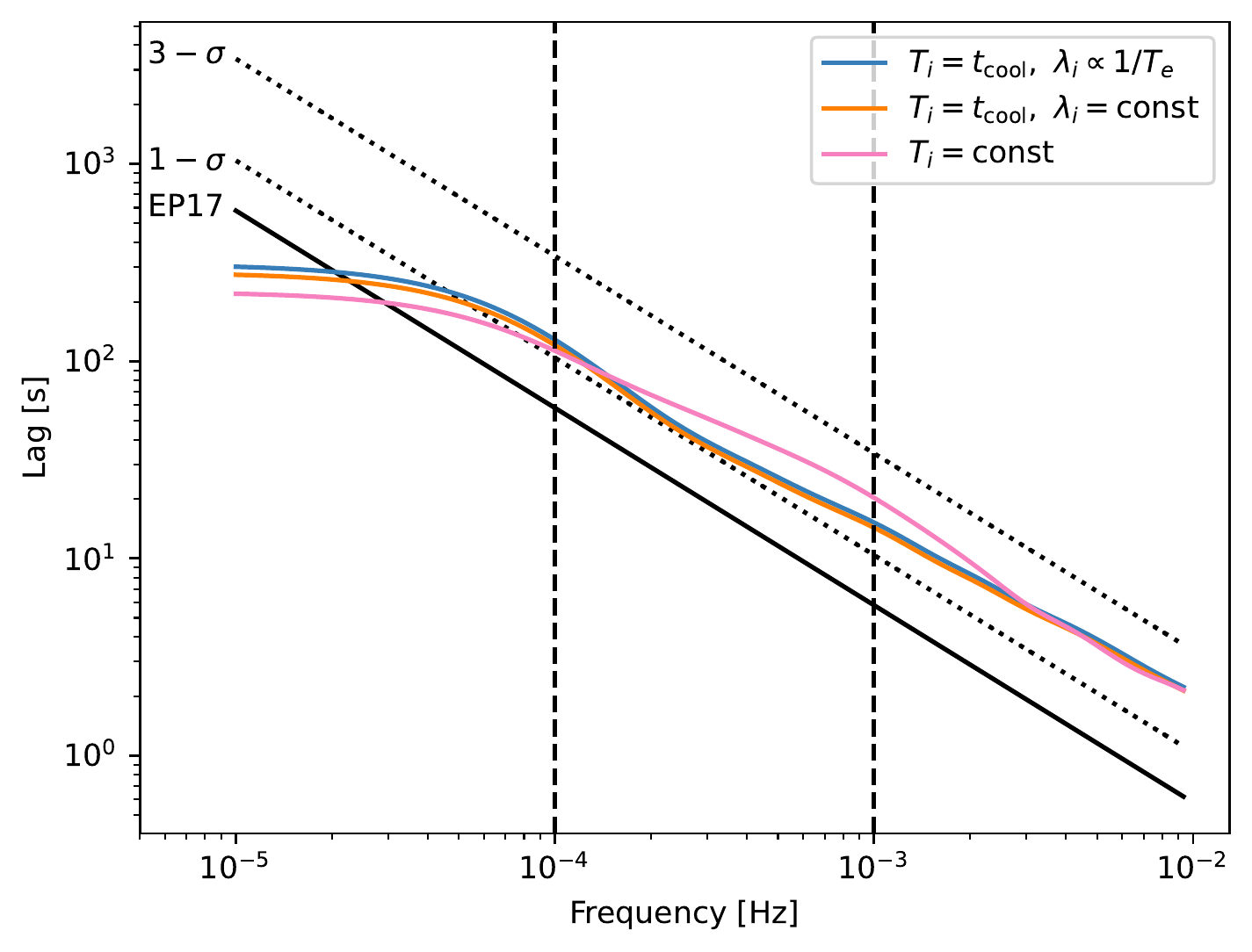}
 \caption{The time-lags of multiple coronae whose duration and probability depends on the physical properties of the coronae. Blue:  $T_i=t_{\rm cool,i}$, and $\lambda_i\propto kT_e,i^{-1}$. Orange: $T_i=t_{\rm cool,i}$, but $\lambda$ is constant. The solid line with pink color is identical to the same color curve in Fig.~\ref{fig:multi_dist}. The vertical dashed lines indicates the frequency range $10^{-4}-10^{-3}~\rm Hz$.}
 \label{fig:multi_shot}
\end{figure}

The results when taking the duration of each flare to be $t_{\rm cool}$ are presented in Fig.~\ref{fig:multi_shot}. The parameters are the same as the case of uniform distribution of $h$ and $\Rc$ in Fig.~\ref{fig:multi_dist}, except that here we assume $h_{\rm max}=100~\Rg$. The blue solid line indicates the time-lags in the case when  the life-time of the corona is its cooling timescale, $T_i=t_{{\rm cool,}i}$, and $\lambda_i\propto kT_{e,i}^{-1}$. This assumption implies that a hotter corona is less likely to appear. As the cooling time-scale is almost inversely proportional to the corona temperature, this assumption also implies that a short-lived corona is less frequent. The orange line shows the time-lags when the corona life-time is again equal to $t_{{\rm cool},i}$, but the probability of a corona appearing is the same in all of them. The solid, pink line is like the same color line in Fig.~\ref{fig:multi_dist} (except that $h_{\rm max}=100~\Rg$ here), and we plot it for comparison reasons. This line breaks to a roughly constant time-lag at a lower frequency when compared with the respective model in  Fig.~\ref{fig:multi_dist}, because $h_{\rm max}$ (and hence $R_{c,{\rm max}}$ as well) is larger now.

The model time-lags are similar in all cases, which implies that the corona life-time does not affect significantly the time-lags shape. Again, this is because the time lag at each frequency is mainly determined by the coronae with $\nu_{0r}\approx \nu_{0i} \approx \nu$. Therefore, the sinc term in eq.~\ref{eq:shotcs_multi} is cancelled out when we divide the imaginary over the real part to compute the time-lags. It is apparent that all three curves more or less resemble the profile of the observed time-lags, over a broad frequency range of $\sim 5\times 10^{-5}-10^{-2}~\rm Hz$. The amplitude of the time-lags agree with the observations within $1-2~\sigma$.

Given that the time-lags are insensitive to the temperature of the seed photons, we do not expect the results to change much with the temperature of the seed photons. As the time-lags are nearly linearly dependent on $M_{\rm BH}$, we expect that, for an AGN with a different black hole mass, the same Compton time-lags (i.e. time-lags which bend to a constant value below the same low frequency) can be obtained by only changing $h_{\rm max}$.

\section{Summary and discussion}

We present the results from a systematic investigation of the Fourier time-lags due to Comptonization in a spherical, homogeneous and isothermal X--ray corona, which is located on the rotational axis of the central BH. The calculations are performed with \codename{}, a GR Monte-Carlo radiative transfer code. We do not assume mono-energetic seed photons. Instead, we consider photons emitted by an accretion disc with a Novikov-Thorne temperature profile, and we compute exactly their direction and energy as they enter the corona. Due to its Monte-Carlo nature, we are able to precisely treat the scattering process, and we can model the response of the disc-corona system, assuming a broad range of values for the corona physical properties (i.e.\ radius, height, temperature, and optical depth), various accretion rates, and BH masses. \codename{} also enables us to assess the effect of the strong gravitational field of the black hole on the lag spectra, which is important when the corona is close to the black hole.

We do not consider extended coronae with a density gradient \citep[like the model of][]{hua_phase_1997,kazanas_temporal_1997} or accreting coronae with additional physical processes (e.g.\ inward propagating fluctuations) as it is done in recent works \citep[e.g.][]{mahmoud_physical_2018,chainakun_x-ray_2019}. Our objective is to study the constraints on the corona parameters due to Comptonization, only. As long as thermal Comptonization is the physical mechanism that produces the X-rays in AGNs, the time-lags we study (under the assumed geometry) should always be present, and they should be considered when a particular variability process is assumed. We provide an example as to how this can be accomplished in practise, by studying the time-lags in the case of a dynamic corona, whose size, height and temperature are variable. 

\subsection{Summary of the results}
\begin{itemize}
\item{The Comptonization time-lags in Fourier domain between energy bands E$_1$ and E$_2$ are well fitted by eq.~\ref{eq:model}: they are constant up to a characteristic frequency, $\nubtl$, and then they rapidly decrease to zero at higher frequencies. The time-lags depend on the difference between the average time required for the seed photon to get up-scattered to E$_1$ and E$_2$, while $\nubtl$ is set by the inverse of the mean photon travel time between subsequent scatterings.} \item{Equations~\ref{eq:tlampall} and \ref{eq:nubreakall} give the time-lags due to thermal Comptonization in the case of a single corona, which is located on the BH rotational axis. We do not expect our results to be sensitive to the actual shape of the corona, as long as the corona major axis is not too larger than its minor axis. Therefore, the equations for the time-lags amplitude and break frequency should be valid for any hot medium, located on top of the BH, and they can be used to get an idea of the delays Comptonization itself will add to the X--ray variations at different energy bands.}
\item{To compute model time-lags (which can be compared with observations), a model for the variability of the X--ray properties of the corona must be considered. To take into account the time-delays due to Comptonization, the model should be formulated in terms of a random series of X--ray coronae with variable physical parameters. As long as the corona changes its properties randomly, without any correlation between the time when the coronae appear and their physical properties, eq.~\ref{eq:shotcs_multi}, together with eqs.~\ref{eq:modelreal} -- \ref{eq:modelim} and eqs.\,\ref{eq:cross_amp_real}--\ref{eq:ai}, can be used to compute the  cross-spectrum, as well as the time-lags.}
\item{We also provide an equation for the (observed) 2--10 keV X--ray luminosity and the cooling time scale of a single, spherical corona, as a function of the its physical characteristics and its location above the BH (eq.~\ref{eq:lambdax} and \ref{eq:tcooleq}, respectively).}
\end{itemize}

\subsection{The importance of the general relativistic effects}

We expect the shape of the time-lags, as determined by eq.\,\ref{eq:model}, to be the same irrespective of whether the X--ray corona is close or further away from the central BH. However, we do expect the time-lags parameters, i.e. the time-lags amplitude and break frequency to be affected when GR effects are strong. Indeed, the results we obtain in this work show that GR effects play a role in determining these parameters, especially when the corona is close to the black hole. This is obvious by looking at eqs.~\ref{eq:tlampall} and \ref{eq:nubreakall}, and  Fig.~\ref{fig:heightres}. The time lags-amplitude increase exponentially with decreasing source height. Therefore, the GR effects become important as long as the corona height is smaller than $\sim 5-10$ R$_g$.  The time-lags characteristic frequency is also decreasing with decreasing height, but the dependence on $h$ is linear, so the GR effects are not so strong in this case. \color{black} The reason as to why the time-lags are affected by the GR effects can be found in \S \ref{sec:height}.

GR effects should also affect the time-lags indirectly by changing the energy of the seed photons as they travel from the accretion disc to the corona, since the time-lags amplitude (mainly) also depend on the seed photon energy (\S \ref{sec:seeden}). It is not straightforward to quantify how important the GR effects are in this case, as we must quantify the difference between the mean photon energy entering the corona in the case we omit and when we consider GR, but this is outside the scope of the present work.

In general, we believe at  GR effects cannot be neglected while modeling the Compton time-lags of AGNs, specially when the X-ray corona is located close to the BH. In any case, the results we present in this work do include the GR effects, therefore their use in practice would imply the GR effects are properly taken into account.

\subsection{Comparison with observations}
Comptonization time-lags from a single corona cannot explain the observed time-lags in AGNs, which have a power-law like shape. Nevertheless, if X-rays in AGNs are produced by thermal Comptonization of the disc photons, the model time-lags should be at least consistent with the current observations\footnote{`Consistent' in the sense that we discussed in \S \ref{sec:single_corona}}. We find that under the assumption of a single corona, the model-lags are marginally consistent with the current observations if the corona is very small in sizes and/or the corona has a temperature higher than $\sim 100-200~\rm keV$.

The situation improves significantly when we consider a dynamic corona, whose size and height vary (randomly), and its life-time is equal to the cooling time-scale. As long as the height of the corona can be up to $100~\Rg$, the time-lags, due to Comptonization only, from such a variable corona can have a power-law shape, over a broad range of frequencies. This will always be the case, if there are multiple coronae with different radii. The low frequency limit (below which the time-lags flatten), depends on the largest radius (permitted by the highest height of the corona above the BH). To explain the observed time-lags which follow a power-law form with a slope of $\sim -1$ down to $10^{-4}$ Hz, $h_{\rm max} \sim 100~\Rg$ (and $R_c \sim h_{\rm max}-1$) for a 10$^7$ M$_{\odot}$ BH. The time-lags should decrease exponentially above the frequency $\nu_{0r,i}$ that is set by the smallest corona radius. We assumed a lowest radius of $1~\Rg$, but smaller coronae should extend the power-law form of the time-lags to even higher frequencies. This result is irrespective of whether the average rate of occurrence is constant or depends on the corona properties (like its temperature).

We find that the observed time-lags cannot put constraints on the life time of the corona. However, the life-time, $T_i$, as well as the average rate, $\lambda_i$, should affect the average luminosity of the source. In addition, they should affect the amplitude of the cross-spectrum and, hence, the coherence function. The constraints from the average luminosity and the coherence should determine the allowed distribution of $T_i$ and $\lambda_i$. Such a calculation though is beyond the scope of the present work. 

The similarity between the observations and the model time-lags (within 1--2 $\sigma$) is quite impressive, given the simplicity of the assumed model. The model time-lags are larger than the observed ones, but this may be due to the fact that we have not considered the reverberation time-lags, which should work in the opposite direction. The study of the effects of the time-lags due to X--ray reverberation is beyond the scope of the present work but, in general, the reverberation time-lags will reduce the amplitude of continuum time-lags plotted in Fig.~\ref{fig:multi_shot}, bringing them closer to the observations.

The corona may move along the vertical direction with relativistic velocity, driven by, e.g.\ Compton \citep{beloborodov_plasma_1999,ghisellini_compton_2010}, or magnetic field pressure \citep{merloni_coronal_2002,hawley_magnetically_2006,tchekhovskoy_three-dimensional_2016}. We would tentatively expect the Comptonization time-lags, as observed by a distant observer, to have a smaller amplitude and higher break frequency as compared with the values measured in the rest frame of the corona, due to the special relativistic time dilation:
\begin{equation}
A / A^\prime = \nu_{\rm btl}^\prime / \nu_{\rm btl} = \Gamma (1-\beta {\rm cos}\theta),
\end{equation}
where the prime sign denotes values measured in the rest frame, $\Gamma$ is the Lorentz factor of the corona, $\beta=v/c$ where $v$ is the velocity of the relativistic motion, and $\theta$ is the observer's inclination. 
We note that $A\times \nu_b = {\rm const}$, as $\Gamma$ changes both the amplitude and the break frequency, but the product of the two remains constant. This behavior is similar with changing $\Rc$. We therefore expect that for multiple coronae, changing $\Gamma$ would have the same effect with changing $\Rc$ (see Fig.~\ref{fig:multi_uniform}): only the break frequency would change, but the amplitude would remain the same.

It is also possible that the coronae are moving towards the disc, due to e.g.\ the gravitational field of the black hole. In this case, we would expect the time-lags from multiple coronae to have lower break frequencies in the distant observer's frame as compared to the rest frame of the coronae. This means that we do not need a maximum height as large as $100~\Rg$ to have a powerlaw profile down to a few times $10^{-5}~\rm Hz$ if the coronae are moving towards the disc.

We have studied the time-lags due to Comptonization for an a X-ray corona that is located on the BH rotational axis. Our results imply that a single, stationary corona is not consistent with the observed time-lags in AGNs. On the other hand, a dynamic, corona, with a variable location, size and life-time, can result in time-lags with a power-law shape, which are quite similar to the observed ones. Although such a simple model cannot explain the observations in full, our results demonstrate the importance of time-lags due to thermal Comptonization. These time-lags should be taken into account when variability models for the X-ray time-lags are considered.

We provide equations for the cross-spectrum of the X-ray corona in the 2--4 vs 4--6 keV band. We do not study the time-lags in the other energy bands. However, eqs.~\ref{eq:cross_amp_real} - \ref{eq:ai} can still be used to compute the cross-spectrum between energy bands $E_{i}$ and $E_{\rm ref}$, as long as we multiply $A_{0i}$ by $\log(E_i/E_{\rm ref})/0.23$, to take into account the dependence of the time-lags on energy. We note that the time-lags break frequency depend on $E_i$ and $E_{\rm ref}$ as well (see eq.~\ref{eq:nubreakall}). This implies that $\nu_{0i}$ and $\nu_{0r}$, as well as the slope of the time-lags as a function of frequency, will also depend on the energy separation between $E_i$ and $E_{\rm ref}$. Eq.~\ref{eq:taufinal} can still be used to compute the resulting cross-spectrum, which should give a good representation of the intrinsic one, but consideration of the energy dependence of $\nu_{0i}$ and $\nu_{0r}$ will be necessary for a  more accurate calculation of the time-lags.

Our results depend on the assumed geometry, and we plan to study the Comptonization time-lags in the case of a flattened corona, located either above the disc, or within the disc inner radius, as well as the case of a spherical corona located at various locations above the disc.

\section*{acknowledgements}
WZ and MD acknowledge financial support provided by the Czech Science Foundation grant 18-00533S. WZ also acknowledges the support by the Strategic Pioneer Program on Space Science, Chinese Academy of Sciences through grant XDA15052100. This work is supported by the project RVO:67985815. This work is also supported by The Ministry of Education, Youth and Sports, Czech Republic, from the Large Infrastructures for Research, Experimental Development and Innovations project ``e-Infrastructure CZ - LM2018140''. WZ would like to thank IA, FORTH for hospitality. IEP would like to thank ASU for hospitality and acknowledges support from the International Space Science Institute (ISSI), Bern. We thank E. Kammoun for helpful discussions. This research makes uses of \textsc{Matplotlib} \citep{hunter_matplotlib:_2007}, a Python 2D plotting library which produces publication quality figures.

\section*{data availability}
The authors agree to make simulation data supporting the results in this paper available upon reasonable request.



\section*{Appendix}
\begin{figure}
    \centering
    \includegraphics[width=\columnwidth]{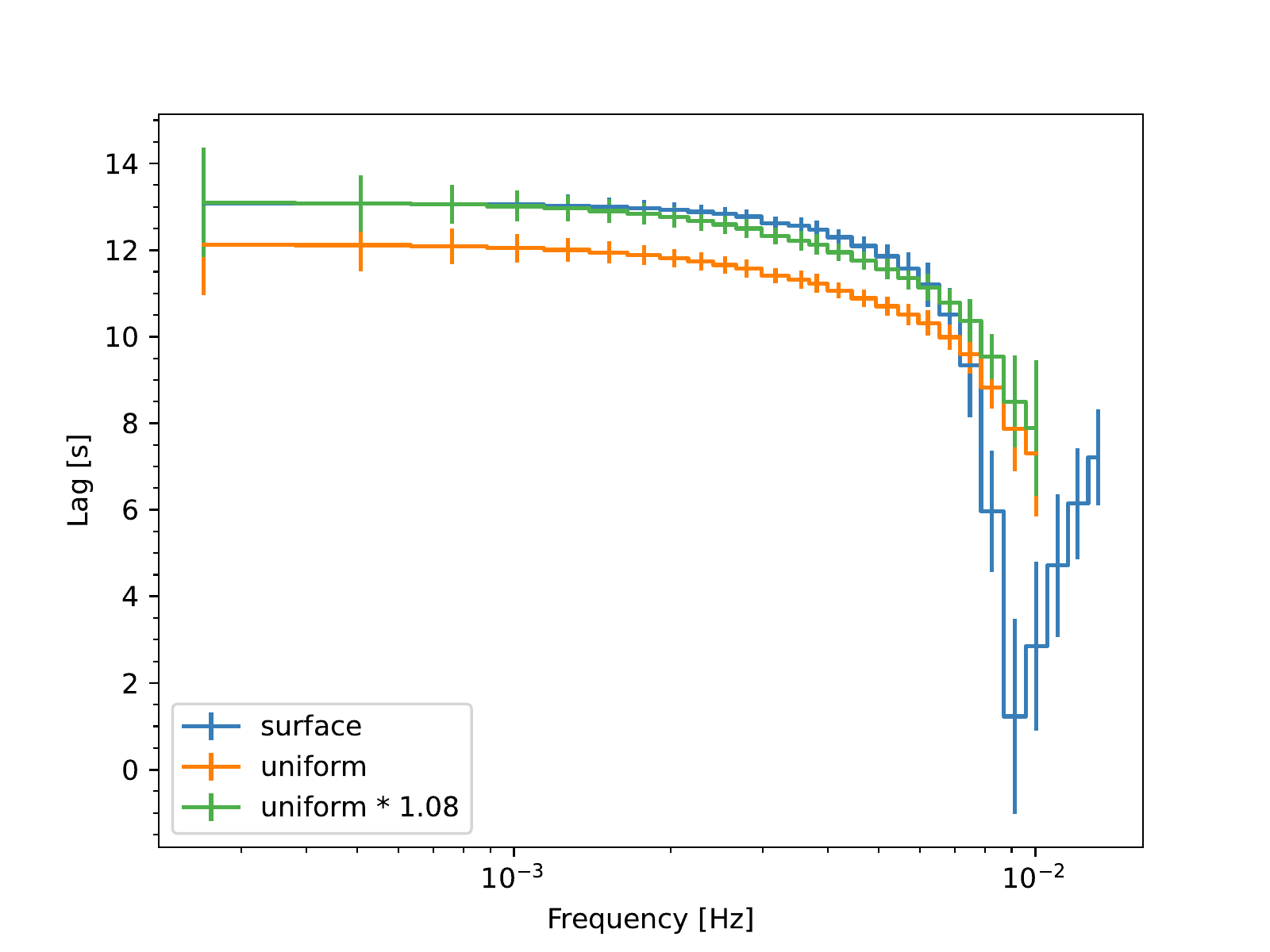}
    \caption{A comparison of time-lags of spherical coronae with different spatial distributions of the seed photons. Blue: the seed photons are located on the surface of the corona. Orange: the seed photons are uniformly distributed in the corona. Green: 1.08 times the time-lags of the uniform case.\label{fig:sr_surface_uniform}}
\end{figure}

We examine if the time-lags due to Comptonization are sensitive to the spatial distribution of the seed photons. For this purpose, we simulate the lightcurves in two cases of the spherical thermal plasma Comptonizing low-energy seed photons while the seed photons have different spatial distributions: in the first case the seed photons are located on the surface of the coronae, and in the other one the seed photons are distributed uniformly in the plasma. We assume $\Te=100~\rm keV$, $\taut=1$, the radius of the plasma to be $1~\Rg$, and the temperature of the seed photons to be $0.01 ~\rm keV$. In Fig.~\ref{fig:sr_surface_uniform} we compare the time-lags of the two cases: the surface case in blue and the uniform case in orange. We find that the time-lags of two cases have similar profiles. The break frequencies are almost identical, while the time-lags amplitude is slightly smaller when the photons are distributed throughout the corona. However, the difference is minimal (less than 10 per cent). This is shown by the fact that the green curve, which is 1.08 times the time-lags of the uniform case, is identical  with the time-lags of the surface case below $\sim 10^{-2}~\rm Hz$.

\label{lastpage}
\end{document}